\documentclass[aps,showpacs]{revtex4}
\usepackage{epsfig}

\def\eqn#1{Eq.~(\ref{#1})}
\def\nn{\nonumber \\}

\def\order#1{{\cal O}\!\left(#1\right)}
\newcommand{\ba}{\begin{eqnarray}}
\newcommand{\ea}{\end{eqnarray}}
\newcommand{\be}{\begin{equation}}
\newcommand{\ee}{\end{equation}}

\begin{document}

\title{
%
%
\[ \vspace{-2cm} \]
\noindent\hfill\hbox{\rm } \vskip 1pt \noindent\hfill\hbox{\rm
Alberta Thy 17-04} \vskip 10pt
Three-loop QCD corrections and $b$-quark decays}

\author{John Paul Archambault}
\altaffiliation{Present address: Department of Physics, Carleton University, Ottawa, ON, Canada K1S 5B6}
\affiliation{
Department of Physics, University of Alberta\\
Edmonton, AB, Canada  T6G 2J1}

\author{Andrzej Czarnecki}
\affiliation{
Department of Physics, University of Alberta\\
Edmonton, AB, Canada  T6G 2J1}

\begin{abstract}
We present three-loop (NNNLO) corrections to the heavy-to-heavy quark transitions in the limit
of equal initial and final quark masses.  In analogy with the previously found NNLO corrections, the
bulk of the result is due to the $\beta_0^2\alpha_s^3$ corrections.  The remaining genuine three-loop
effects for the semileptonic $b\to c$ decays are estimated to increase the decay amplitude by
$0.2(2)\%$. The perturbative series for the heavy-heavy axial current converges very well.
\end{abstract}


\maketitle

\section{Introduction}
The $b$ quark decays most frequently into another relatively heavy
quark, the $c$.  These decays, especially when semileptonic,
provide insight into the properties of
heavy hadrons and an opportunity to determine
Standard Model parameters such as the Cabibbo-Kobayashi-Maskawa
(CKM) matrix element $V_{cb}$ \cite{Uraltsev:2004ta}.

In order to achieve good accuracy in those studies, one needs to
know to what extent gluon emissions modify the weak transition
amplitude.  The evaluation of such effects is complicated by the
presence of three energy scales: the masses of the radiating $b$ and $c$
quarks, and the momentum transfer to the leptons emitted in
the decay process.  The effects of a single gluon (next-to-leading
order, NLO) can be determined as an analytical function of all
three scales (see e.g. \cite{alt,jk2,cj94} and the references cited there),
but in the next-to-next-to-leading order (NNLO), an analytic
expression is known only in the limiting case in which the $c$ quark remains
at rest with respect to the $b$ quark \cite{Czarnecki:1997cf},  known as the zero-recoil
configuration.  It
corresponds to the emission of the charged lepton and the neutrino
in opposite directions.

Beyond the zero-recoil limit, numerous expansions have been
employed to study semileptonic $b$ decays with NNLO accuracy
\cite{MaxRec,Czarnecki:1998kt,Czarnecki:2001cz}.  Those results
improved our knowledge of the $b\to c$ decay rate, the CKM
parameter $V_{cb}$, and in the future may also contribute to the
determination of $V_{ub}$.

\begin{figure}[htb]
\hspace*{25mm}
\begin{center}
\begin{tabular}{c@{\hspace*{10mm}}c@{\hspace*{10mm}}c}
\epsfxsize=40mm \epsfbox{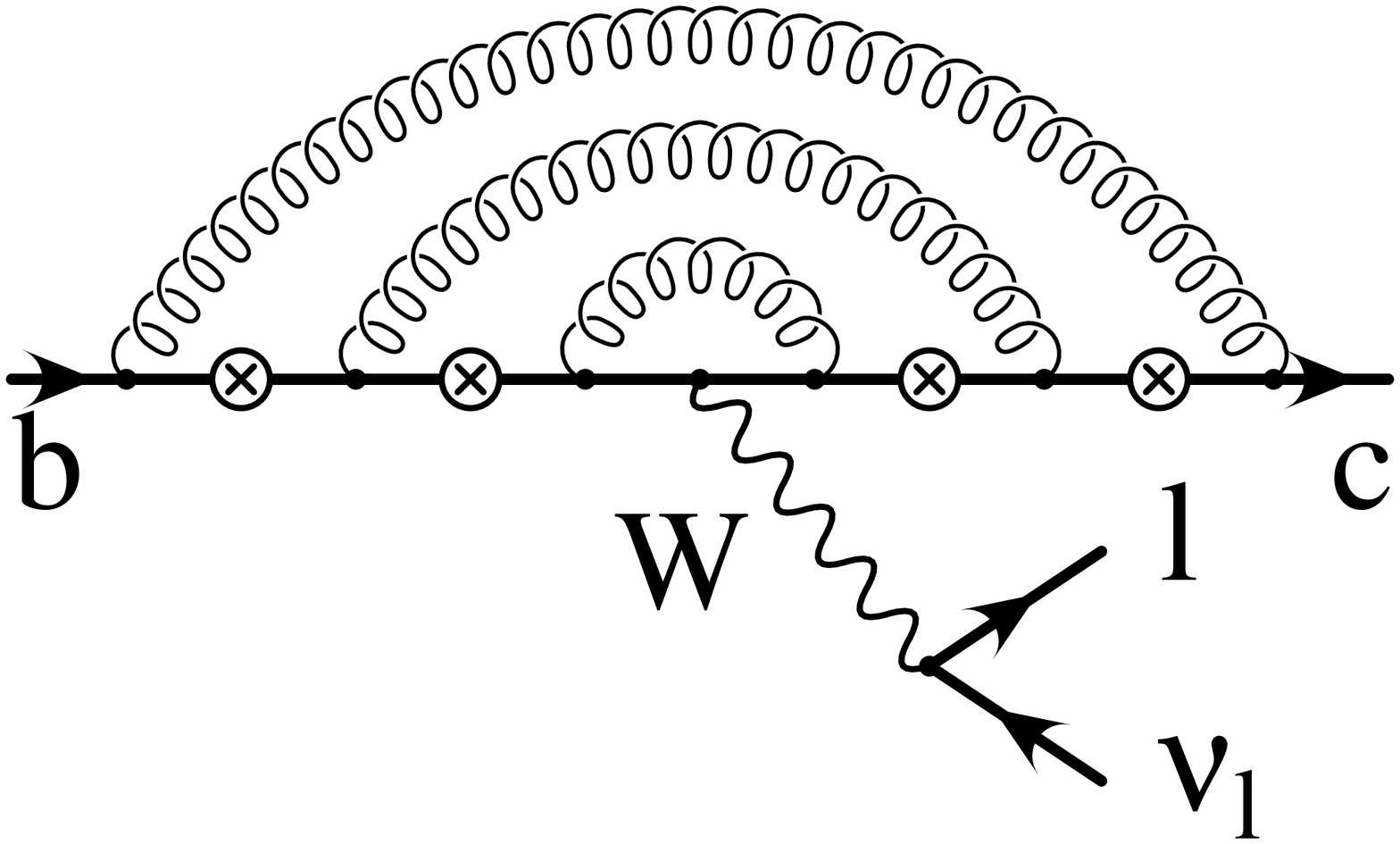} & \epsfxsize=40mm
\epsfbox{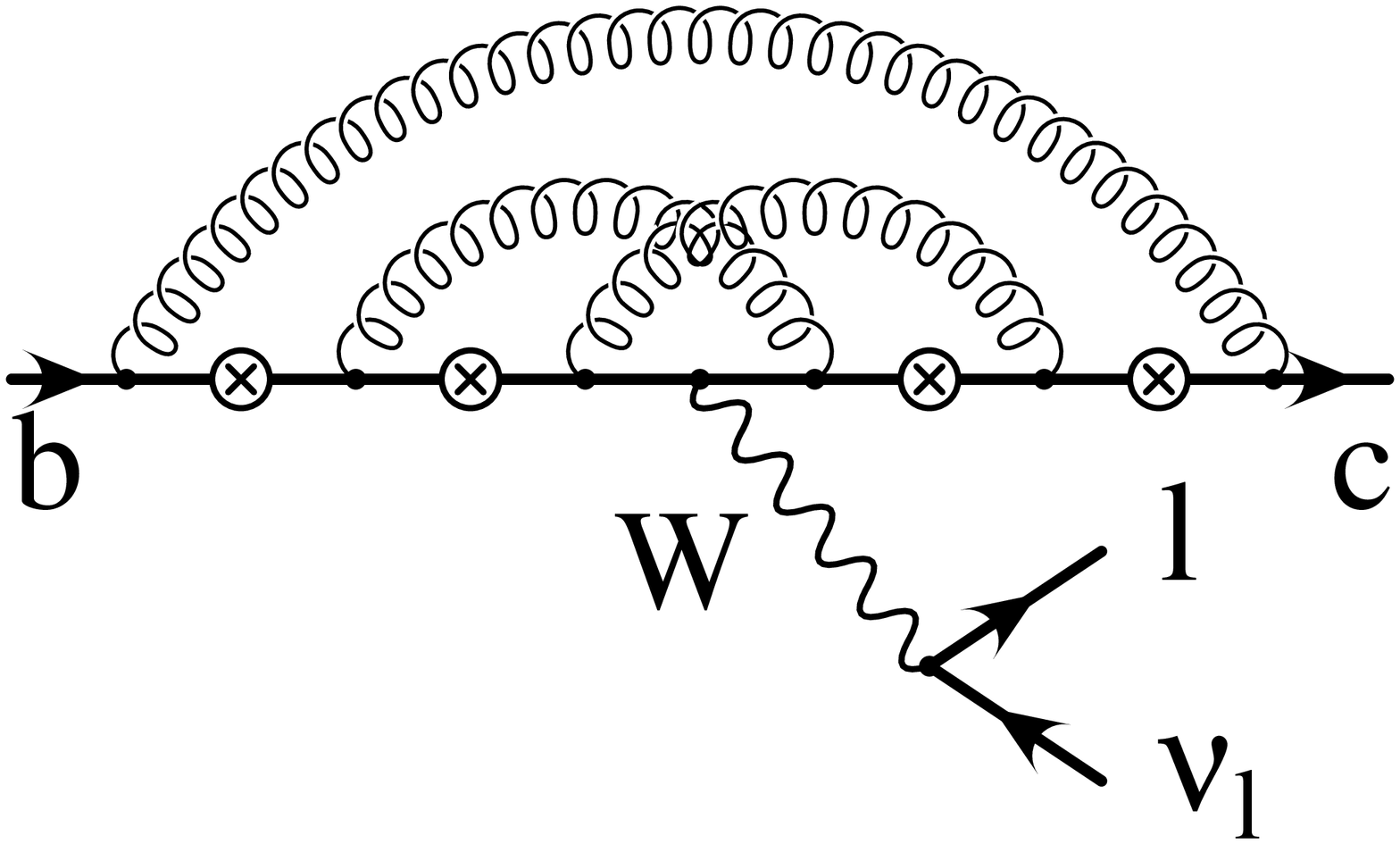} & \epsfxsize=40mm \epsfbox{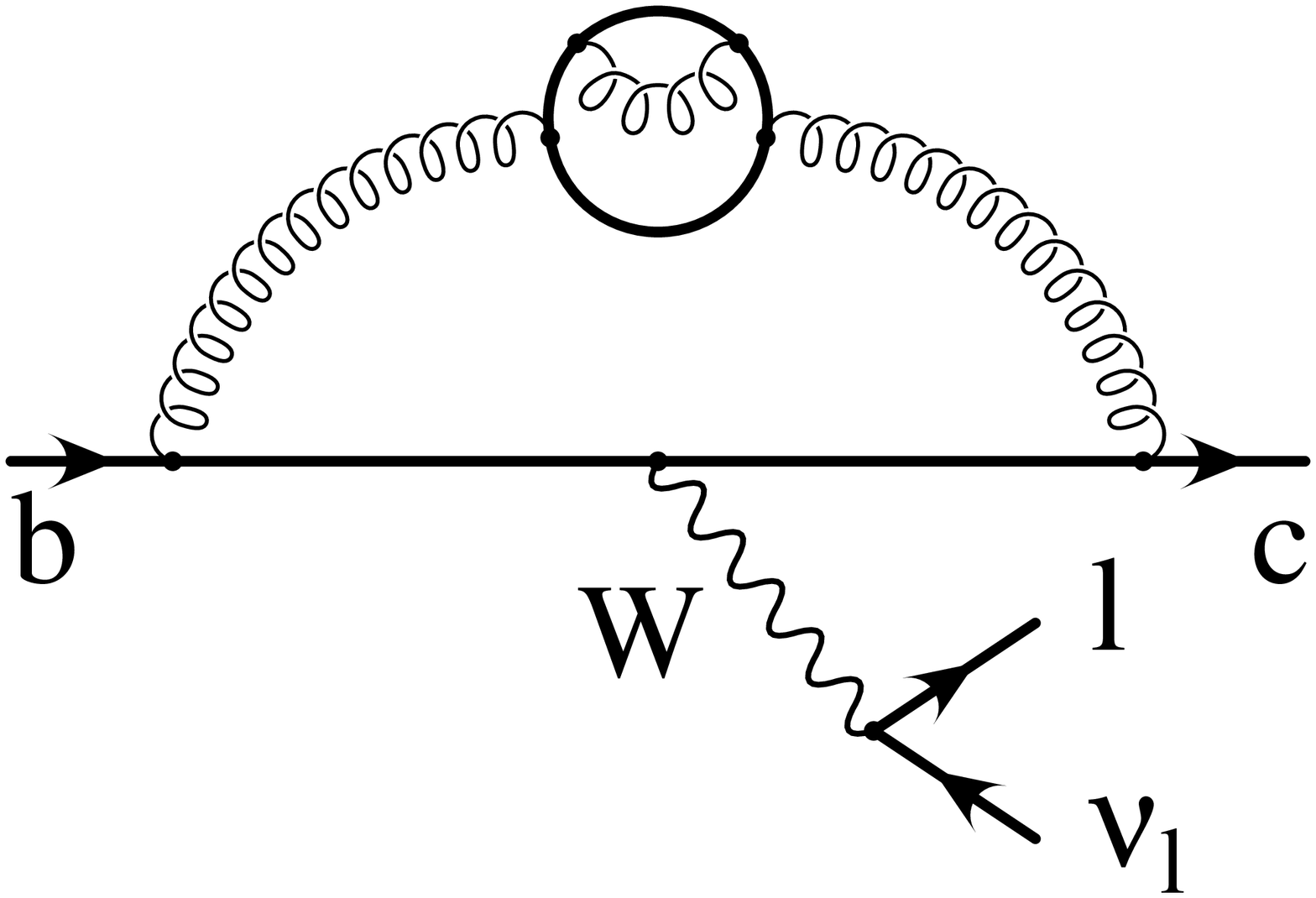}
\end{tabular}
\caption{Examples of three-loop corrections to the semileptonic $b\to c$ decay.
The $W$ can also be emitted from the quark line in the points denoted by crossed circles.
This figure shows only examples of abelian corrections, similar to photonic vertex corrections.
The complete list of such diagrams is given in \cite{ros90}.  Non-abelian diagrams are shown in subsequent
figures.}
\label{fig:3loopsab}
\end{center}
\end{figure}

In the present paper we study effects of the
next-to-next-to-next-to-leading order (NNNLO) to the transition of
one heavy quark into another.  Effects in this order,
$\order{\alpha_s^3}$, have never before been calculated for any
charged particle decay.  The main reason is that the NNNLO
calculations, especially with massive particles in intermediate
states, are very difficult.

Here, we consider a particular kinematical limit in which the
masses of the initial and final state quarks are equal.  This is the
extreme zero-recoil limit: there is no momentum transfer into
leptons and no real gluon radiation.
The zero-recoil
configuration can be realized also with unequal quark masses, when  the leptons, emitted
anti-parallel to each other, carry away all released energy.  Radiative corrections
in such configuration also consist of virtual gluon exchanges only.  If the mass difference
between the quarks is not very large, loop diagrams can be evaluated by expanding in this
mass difference, around the equal mass (``extreme zero-recoil'') limit considered in this paper.

Taking equal quark masses greatly simplifies the
NNNLO calculation because we deal with two-point Green functions,
similar to a quark propagator, with zero-momentum insertions (see
Fig.~\ref{fig:3loopsab}). In fact, the technically related problems of
the three-loop anomalous magnetic moment, Dirac form-factor, and heavy
quark renormalization constants have already been solved
\cite{Laporta:1996mq,Melnikov:1999xp,Melnikov:2000zc}.

For the present project, we modified a program originally written
for the latter two studies \cite{ThanksKirill}.  Thus, with
relatively little effort, we can obtain the first NNNLO
correction to a heavy-to-heavy quark transition and confirm the
expectations of its moderate size. It should be mentioned that
despite the special kinematic limit in which we work, our result
may help estimate NNNLO effects in a variety of other
configurations because most of the kinematic effects reside in
easily  determined phase space effects, while the matrix elements
depend on kinematics only relatively weakly. (There are exceptions to
this picture; for example, along the zero-recoil line, the matrix
elements are affected by a mass-singularity effect, manifesting
itself in logarithms of the initial- and final-state quark mass ratio \cite{Shifman:1987rj,Neubert:1994vy}.
However, the
physical ratio of $b$ and $c$ quark masses is not very large and our result
obtained in the equal mass limit is expected to hold with
reasonable accuracy \cite{Uraltsev95}.)

\section{Corrections to the heavy-to-heavy transition $b\rightarrow cl\nu$}

The effect of QCD corrections on the $W\bar c b$ vertex is parameterized using two functions,
\ba
\gamma_\mu(1-\gamma_5) \to
\gamma_\mu\left[\eta_V(q^2)-\eta_A(q^2)\gamma_5\right].
\ea
In the extreme zero-recoil limit we are considering, the momentum transfer into the $W$ vanishes.
In this case, the vector part of the  interaction does not receive any QCD corrections,
\ba
\eta_V(q^2=0) = 1.
\ea
This is a useful check: the vector coupling part of the sum of all our three-loop diagrams, after
the quark wavefunction renormalization, must vanish.

The axial part of the interaction does receive finite corrections even in the $q^2=0$ limit.
They are expressed
as a power series in $\alpha_s$,
\ba
\eta_A(q^2=0)\equiv \eta_A = 1 + {\alpha_s\over \pi} C_F \eta_A^{(1)}
+ \left( {\alpha_s\over \pi} \right)^2 C_F \eta_A^{(2)}
+ \left( {\alpha_s\over \pi} \right)^3 C_F \eta_A^{(3)} + \order{\alpha_s^4}.
\ea
Throughout this paper, we define $\alpha_s$ in the $\overline{\rm MS}$ scheme at the scale of the quark mass,
and use the pole definition of the quark mass.

\begin{figure}[htb]
\hspace*{25mm}
\begin{center}
\begin{tabular}{c@{\hspace*{10mm}}c@{\hspace*{10mm}}c}
\epsfxsize=40mm \epsfbox{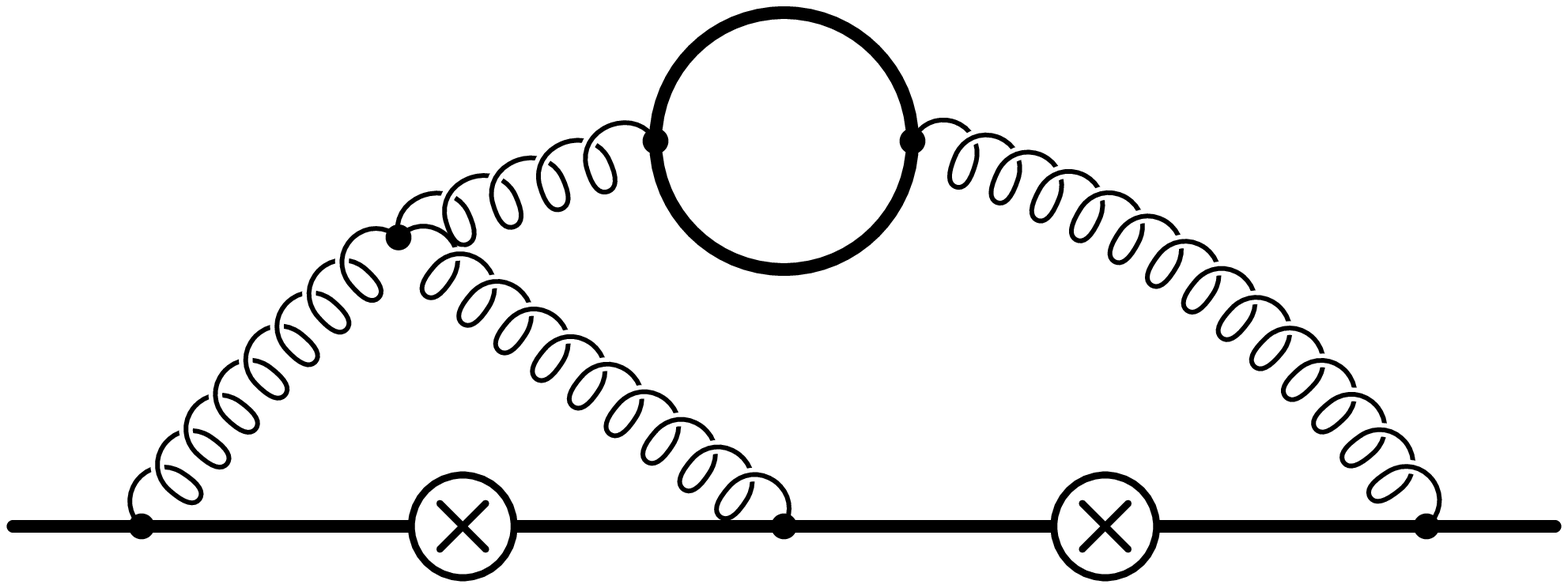} & \epsfxsize=40mm
\epsfbox{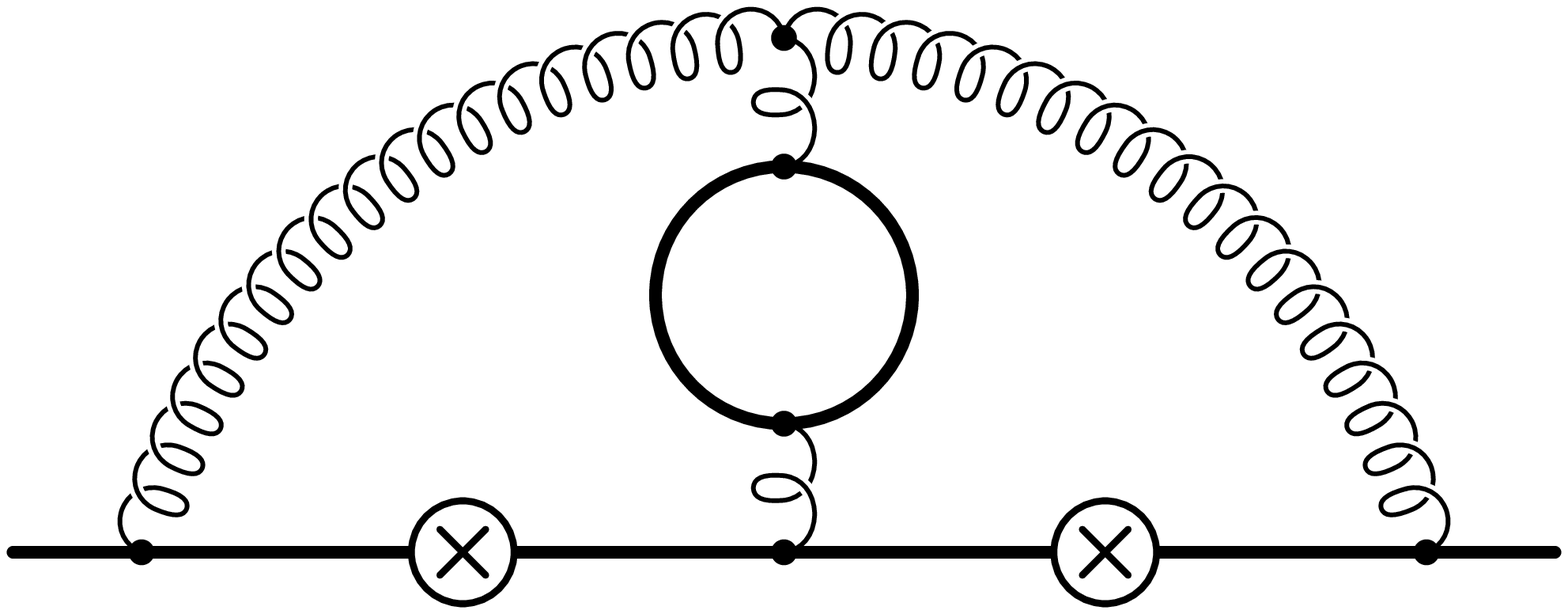} & \epsfxsize=40mm \epsfbox{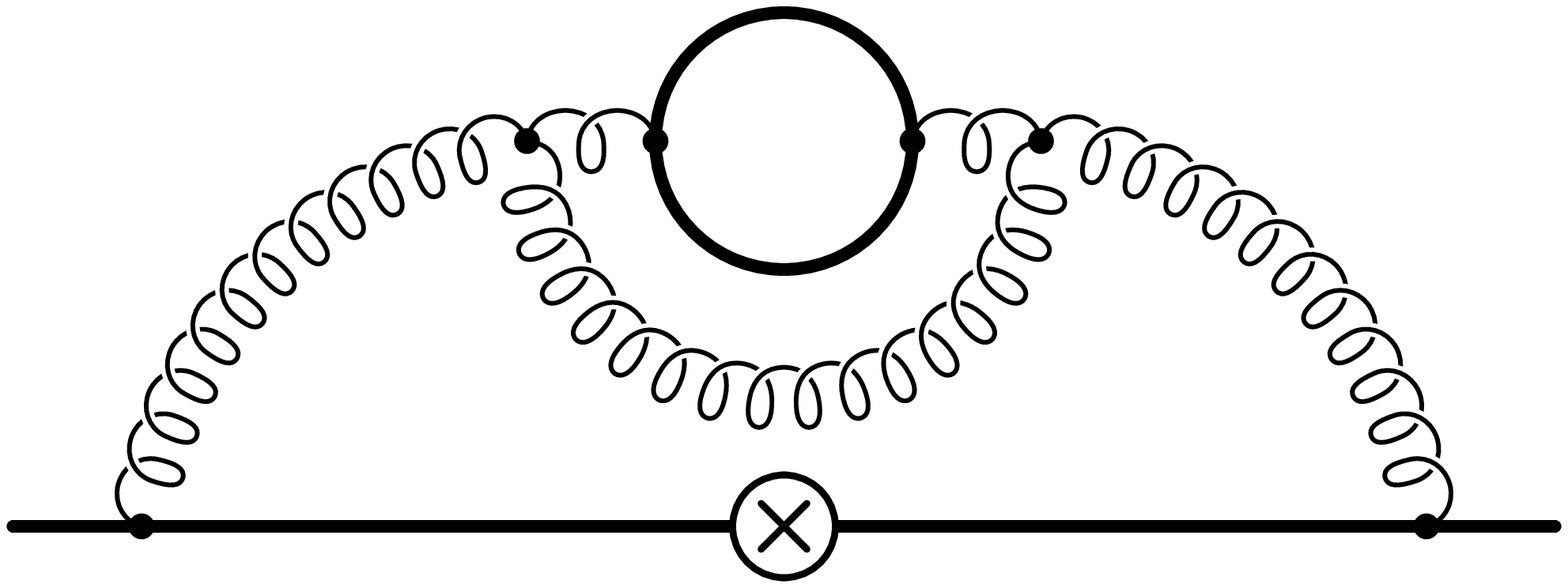}
\\
\epsfxsize=40mm \epsfbox{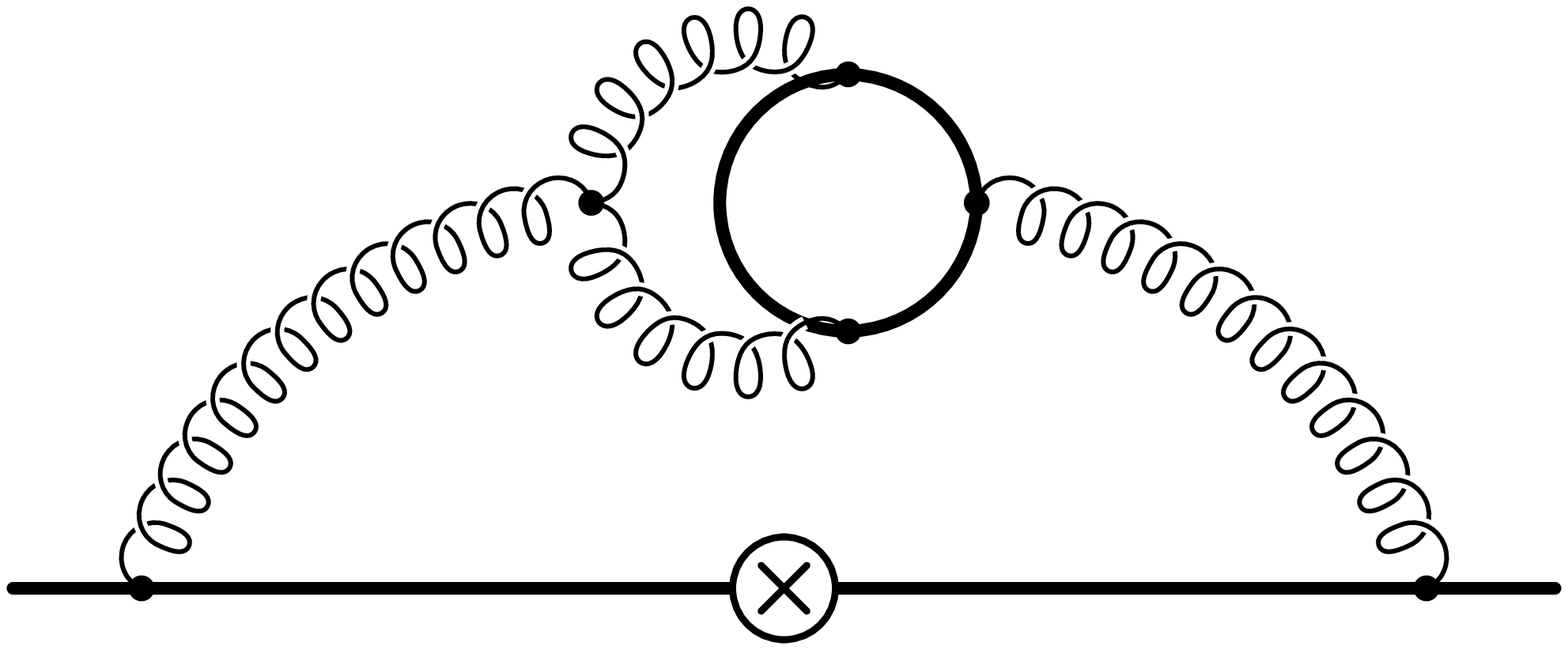} & \epsfxsize=40mm
\epsfbox{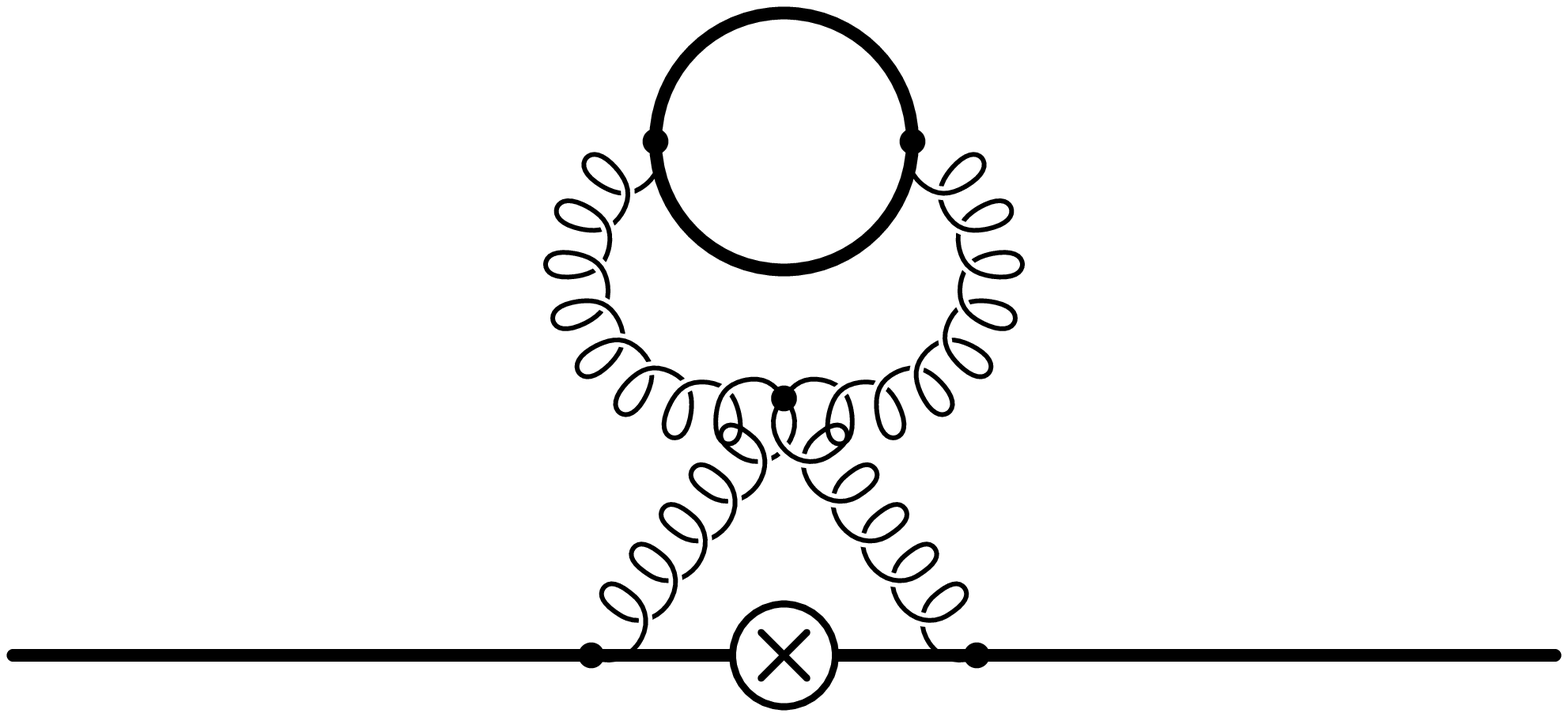} & \epsfxsize=40mm \epsfbox{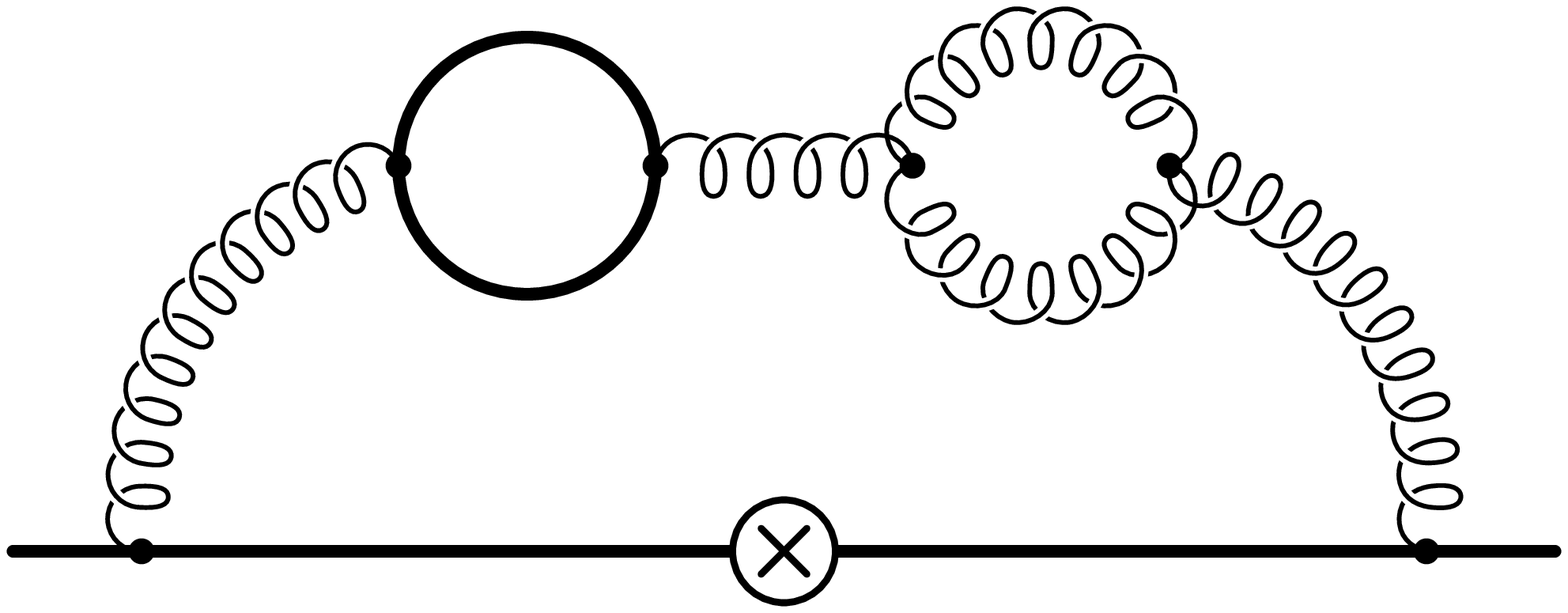}
\end{tabular}
\caption{Three-loop non-abelian diagrams with fermion loops.  Solid line closed loops denote light or
heavy quarks.  As in the previous figure, the $W$ boson can be emitted at any vertex denoted by a crossed circle.}
\label{fig:3loopsnab_NH}
\end{center}
\end{figure}

The first two corrections to $\eta_A$ have been known for a long time
\cite{Shifman:1987rj,Paschalis83,zerorecoil},
\ba
\eta_A^{(1)} &=& -{1\over 2}, \nn
\eta_A^{(2)} &=& C_F\left(-{373\over 144} +{\pi^2\over 6}\right)
+(C_A-2C_F)\left( -{143\over 144} -{\pi^2\over 12} +{\pi^2 \ln 2\over 6} -{1\over 4}\zeta_3\right)
+ N_H T_R \left( {115 \over 36} - {1\over 3}\pi^2\right)
+{7\over 36} N_L T_R  .
\nn
\ea
The SU(3) factors are $C_F=4/3$, $C_A=3$, and $T_R=1/2$, and $N_{H,L}$ denote the numbers of heavy and light
quark flavors.  For $b\to c$ decays, $N_H=2$ and $N_L=3$.  The two- and three-loop corrections
are expressed in terms
of the Riemann zeta function and a tetra-logarithm,
\ba
\zeta_3 &\simeq &  1.202056903,\nn
\zeta_5 &\simeq &  1.036927755, \nn
a_4 &\equiv &{\rm Li}_4\left(1\over 2\right) \simeq   0.5174790617.
\ea
The result of the study reported here is the
third correction, $\eta_A^{(3)}$.  We find
\ba
&& \hspace*{-5mm}\eta_A^{(3)} = N_H N_L T_R^2   \left(  - {425 \over 162} + {7 \over 27} \pi^2 \right)
 \nonumber \\ &&
       + N_H T_R  C_A   \left( {1339 \over 72} + {7 \over 36} \zeta_3 \pi^2 -
       {121 \over 27} \zeta_3
          - {5 \over 4} \zeta_5 + {136 \over 27} \pi^2 \ln 2 + {1 \over 3} \pi^2
 \ln^2 2 - {1019 \over 243} \pi^2
       - {53 \over 1080} \pi^4 - {1 \over 3} \ln^4 2 - 8 a_4 \right)
\nonumber \\ &&       + N_H T_R C_F
  \left( {7679 \over 2592} - {355 \over 108} \zeta_3 +
       {16 \over 27} \pi^2
         \ln 2 - {4 \over 9} \pi^2 \ln^2 2 - {433 \over 486} \pi^2 + {1 \over 30
} \pi^4 + {4 \over 9} \ln^4 2 +
       {32 \over 3} a_4
          \right)
\nonumber \\ &&       + N_H^2 T_R^2   \left(  - {1055 \over 324} + {8 \over 3} \zeta_3 \right)
\nonumber \\ &&       + N_L T_R  C_A   \left( {469 \over 648} + {19 \over 18} \zeta_3 -
       {5 \over 27} \pi^2 \ln 2
          + {2 \over 27} \pi^2 \ln^2 2 + {49 \over 216} \pi^2 - {11 \over 648} \pi^4
           + {1 \over 27} \ln^4 2 +
       {8 \over 9} a_4
          \right)
\nonumber \\ &&       + N_L T_R C_F   \left( {2293 \over 864} - {23 \over 18} \zeta_3 +
       {10 \over 27} \pi^2 \ln 2
          - {4 \over 27} \pi^2 \ln^2 2 - {14 \over 27} \pi^2 + {11 \over 324}
          \pi^4 - {2 \over 27} \ln^4 2 -
       {16 \over 9} a_4
          \right)
\nonumber \\ &&       + N_L^2 T_R^2    \left( {25 \over 324} - {1 \over 27} \pi^2 \right)
\nonumber \\ &&       +  C_A^2   \left(  - {16241 \over 5184} + {11 \over 144} \zeta_3 \pi^2 +
       {215 \over 144}
          \zeta_3 - {5 \over 3} \zeta_5 + {139 \over 108} \pi^2 \ln 2 - {25
          \over 108} \pi^2 \ln^2 2 -
       {1423 \over 1728} \pi^2
          + {97 \over 6480} \pi^4 - {2 \over 27} \ln^4 2 - {16 \over 9} a_4 \right)
\nonumber \\ &&       + C_F C_A   \left(  - {2723 \over 864} + {1 \over 18} \zeta_3 \pi^2 -
       {79 \over 24} \zeta_3 +
         {65 \over 12} \zeta_5 - {215 \over 108} \pi^2 \ln 2 + {10 \over 27} \pi
^2 \ln^2 2 + {467 \over 288} \pi^2
       - {359 \over 6480} \pi^4 + {13 \over 54} \ln^4 2 + {52 \over 9} a_4 \right)
 \nonumber \\ &&      + C_F^2   \left(  - {1141 \over 576} + {7 \over 12} \zeta_3 \pi^2 +
       {40 \over 9} \zeta_3 - {20 \over 3}
          \zeta_5 - {7 \over 6} \pi^2 \ln 2 + {5 \over 27} \pi^2 \ln^2 2 + {155
\over 216} \pi^2 -
       {7 \over 216} \pi^4 - {5 \over 27} \ln^4 2 - {40 \over 9} a_4 \right).
\label{eq:eta3}
       \ea
Among these ten contributions, differing by SU(3) factors, the first seven come from diagrams containing
one or two fermion loops, with light and/or heavy quarks.  Among them, the second and the fifth arise from
non-abelian diagrams with three- or four-gluon vertices, shown in Fig.~\ref{fig:3loopsnab_NH}.
The remaining
five contributions containing fermion loops are analogous to those in QED, with one example shown in
Fig.~\ref{fig:3loopsab}.

\begin{figure}[h!]
\begin{center}
\hspace*{5mm}
\begin{minipage}{16.cm}
\begin{tabular}{c@{\hspace*{10mm}}c@{\hspace*{10mm}}c}
\vspace*{3mm} \psfig{figure=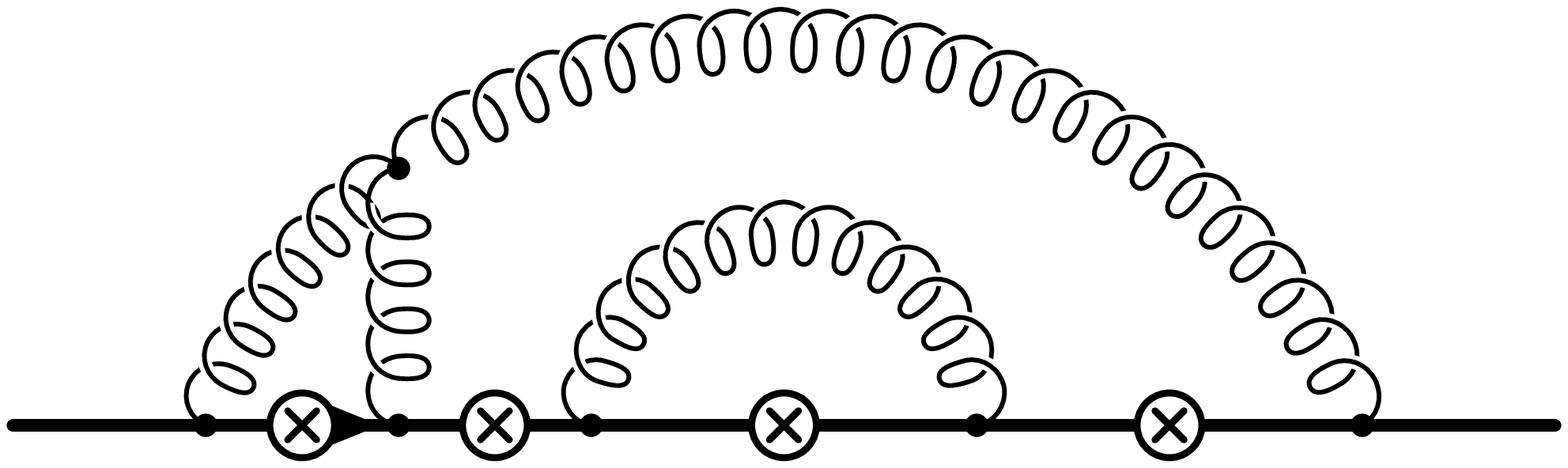,width=40mm} &
\psfig{figure=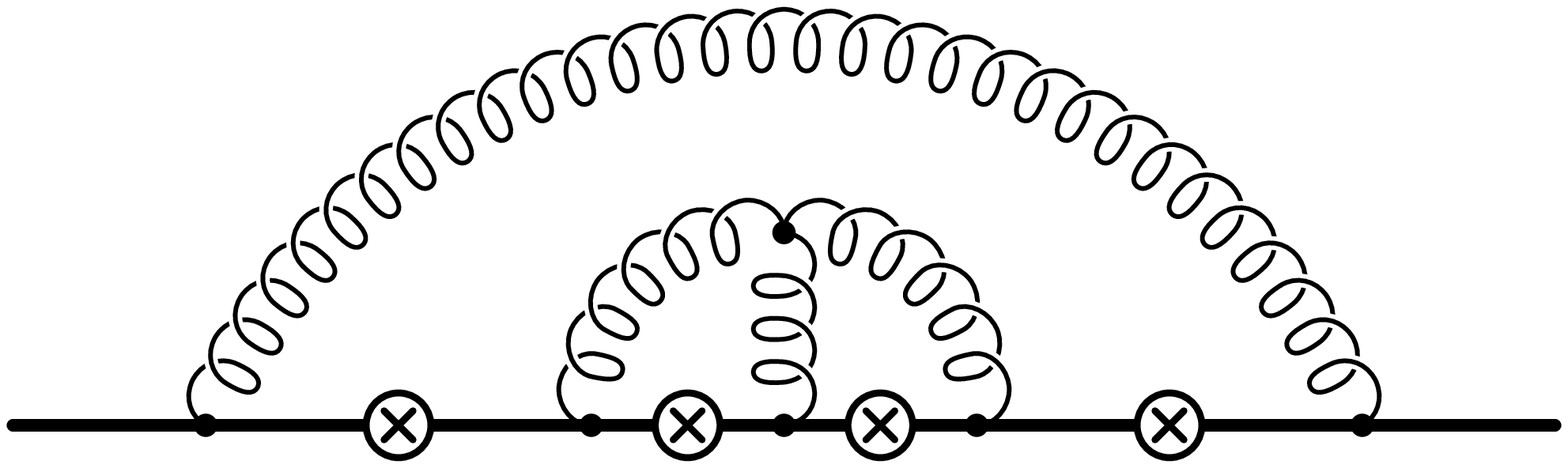,width=40mm} &
\psfig{figure=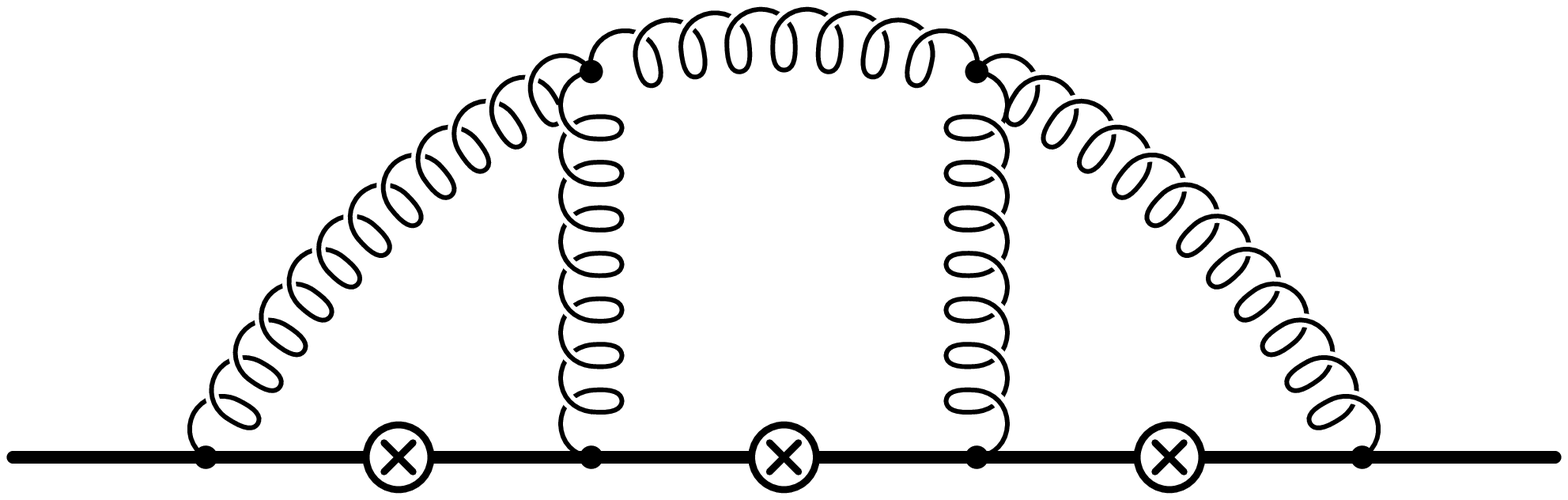,width=40mm}  \\
\psfig{figure=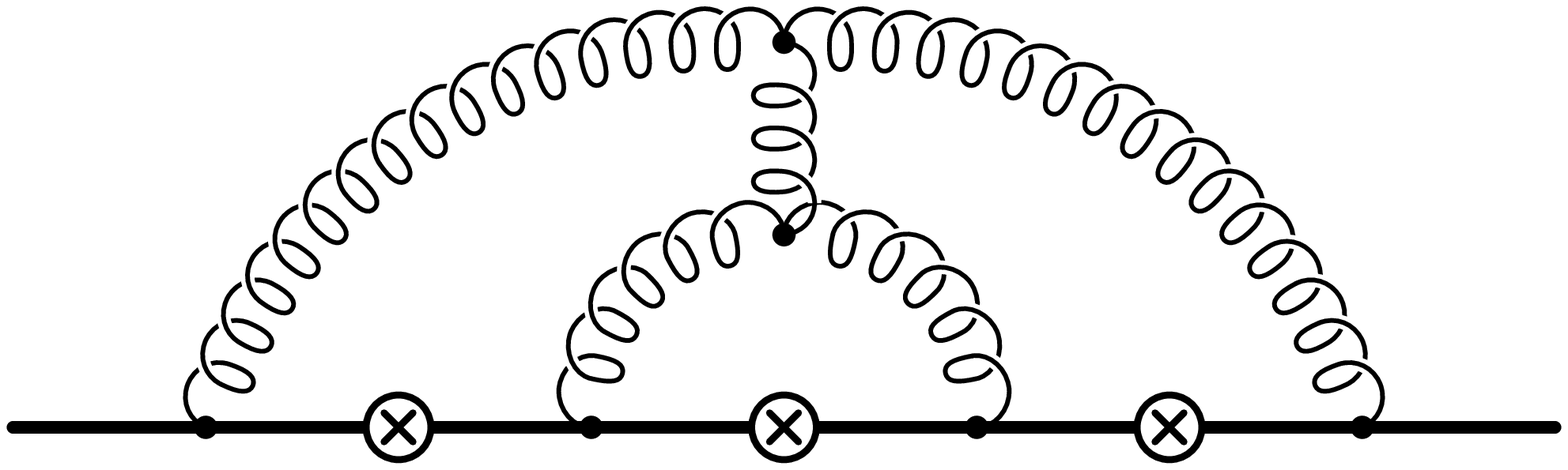,width=40mm}  &
\psfig{figure=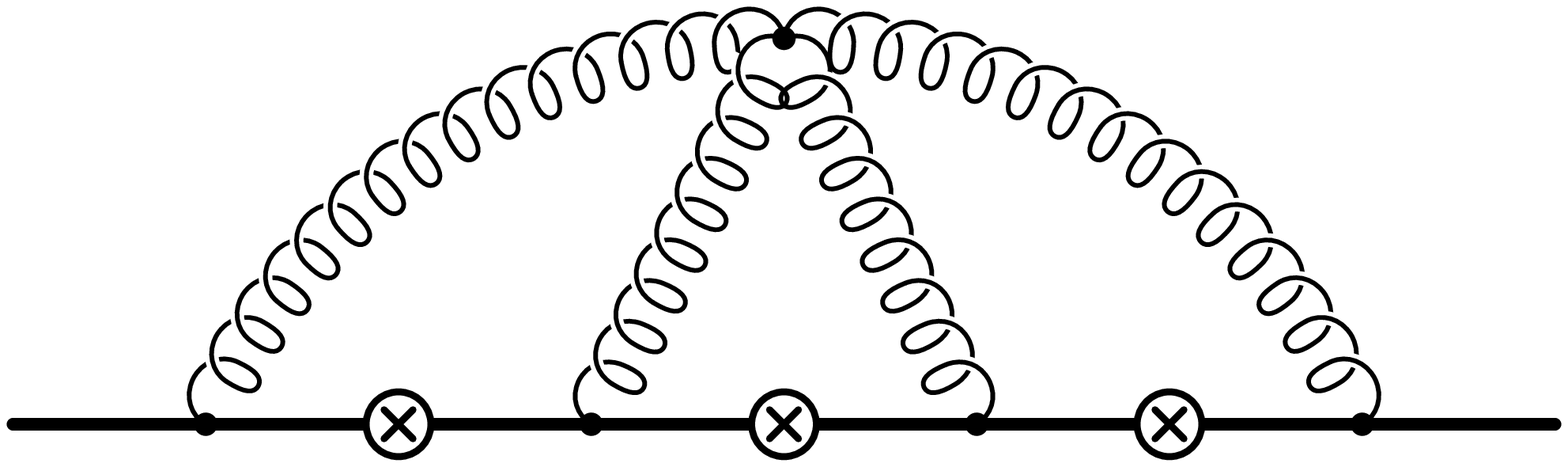,width=40mm}  &
\psfig{figure=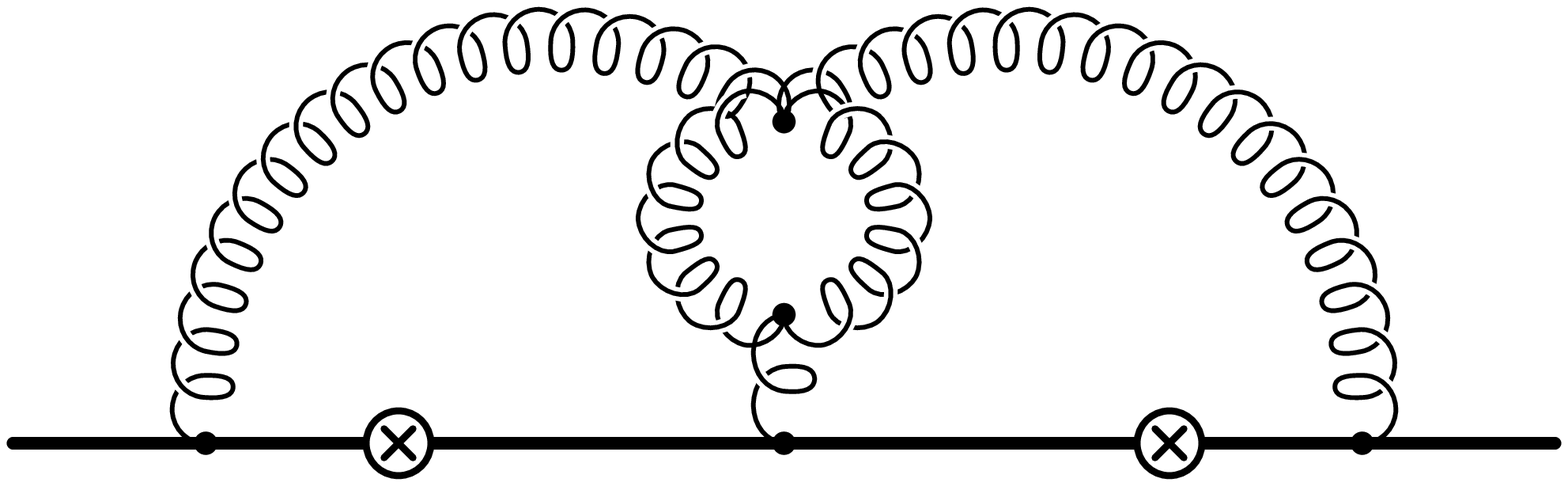,width=40mm}  \\
\psfig{figure=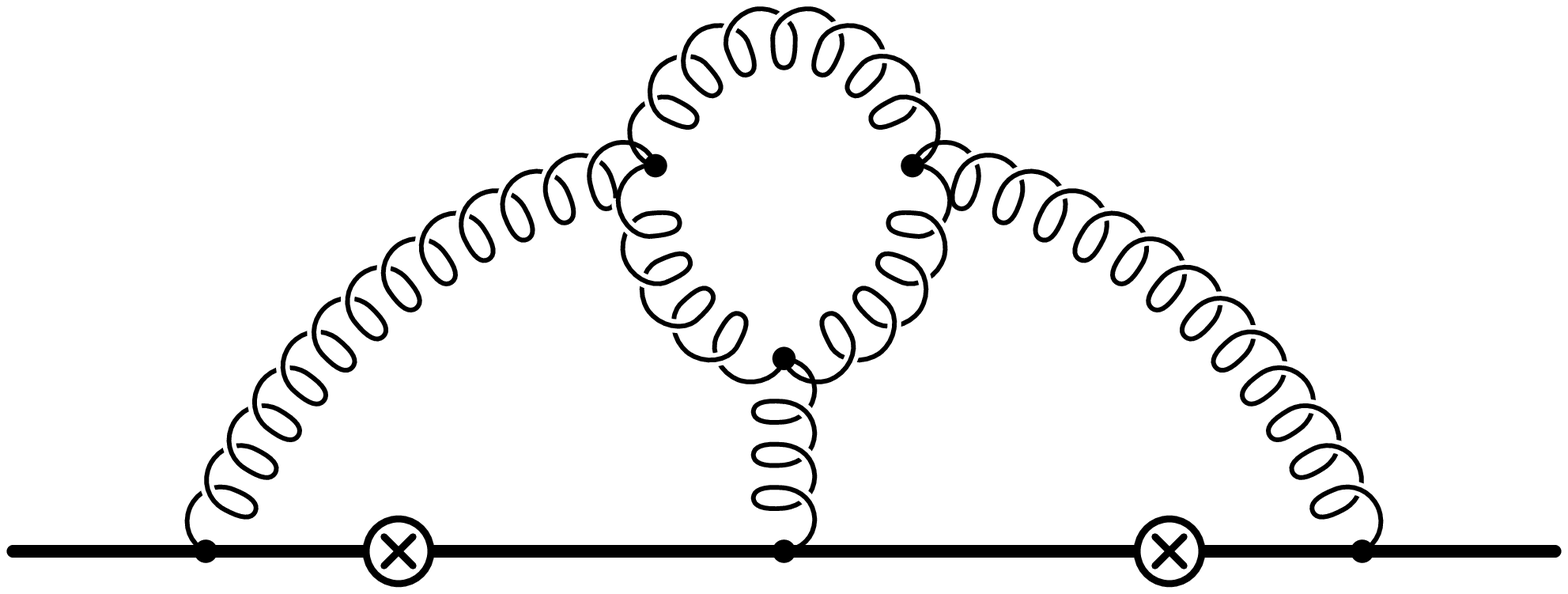,width=40mm}  &
\psfig{figure=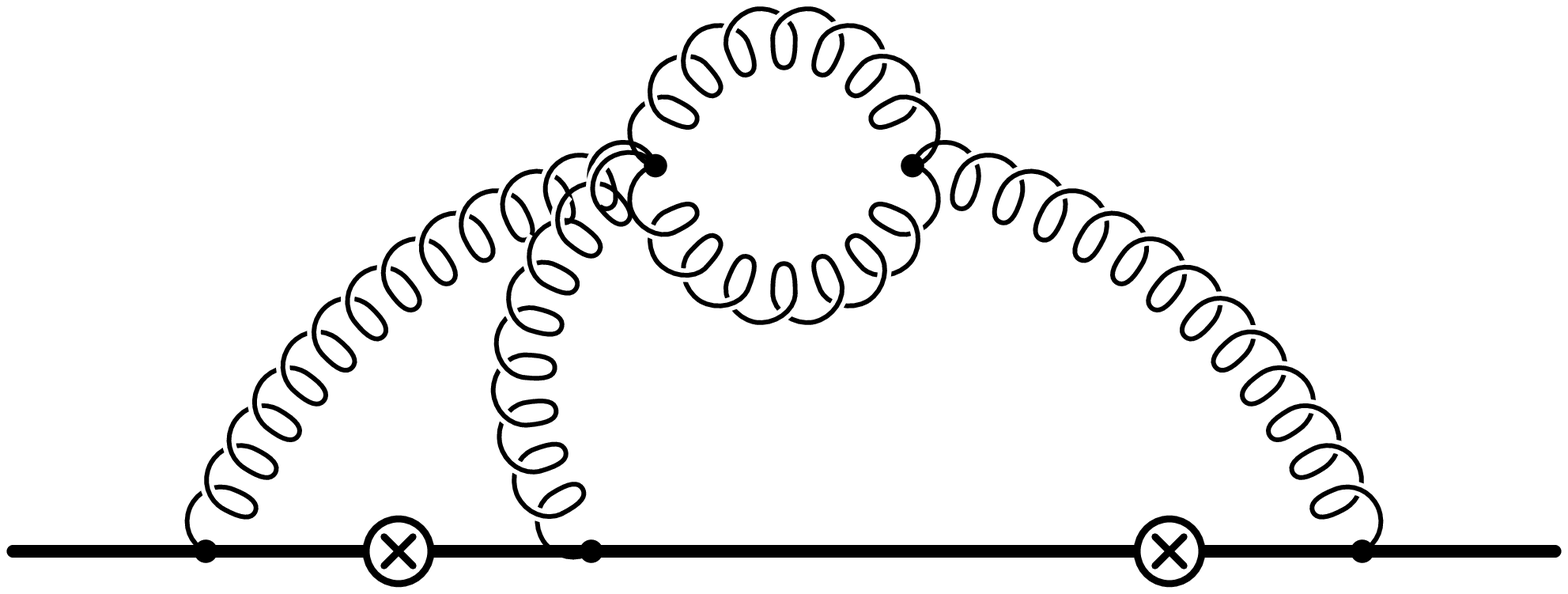,width=40mm}  &
\psfig{figure=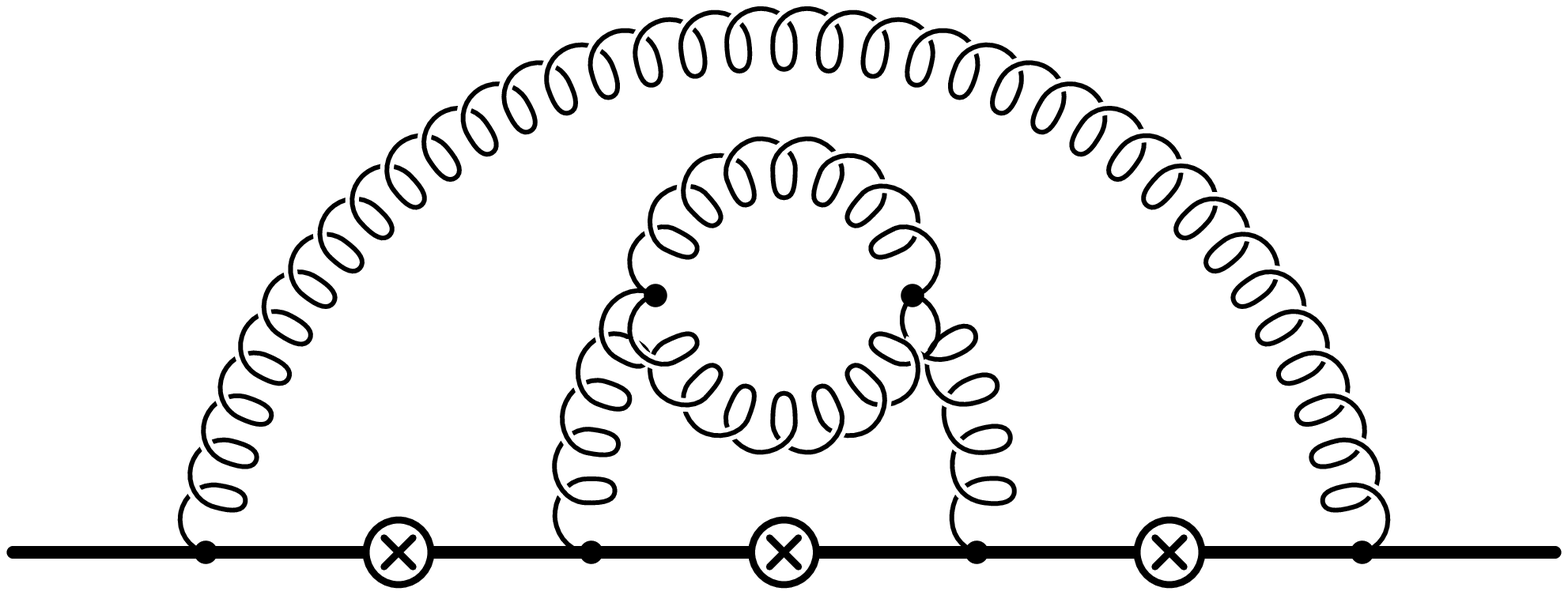,width=40mm} \\
\psfig{figure=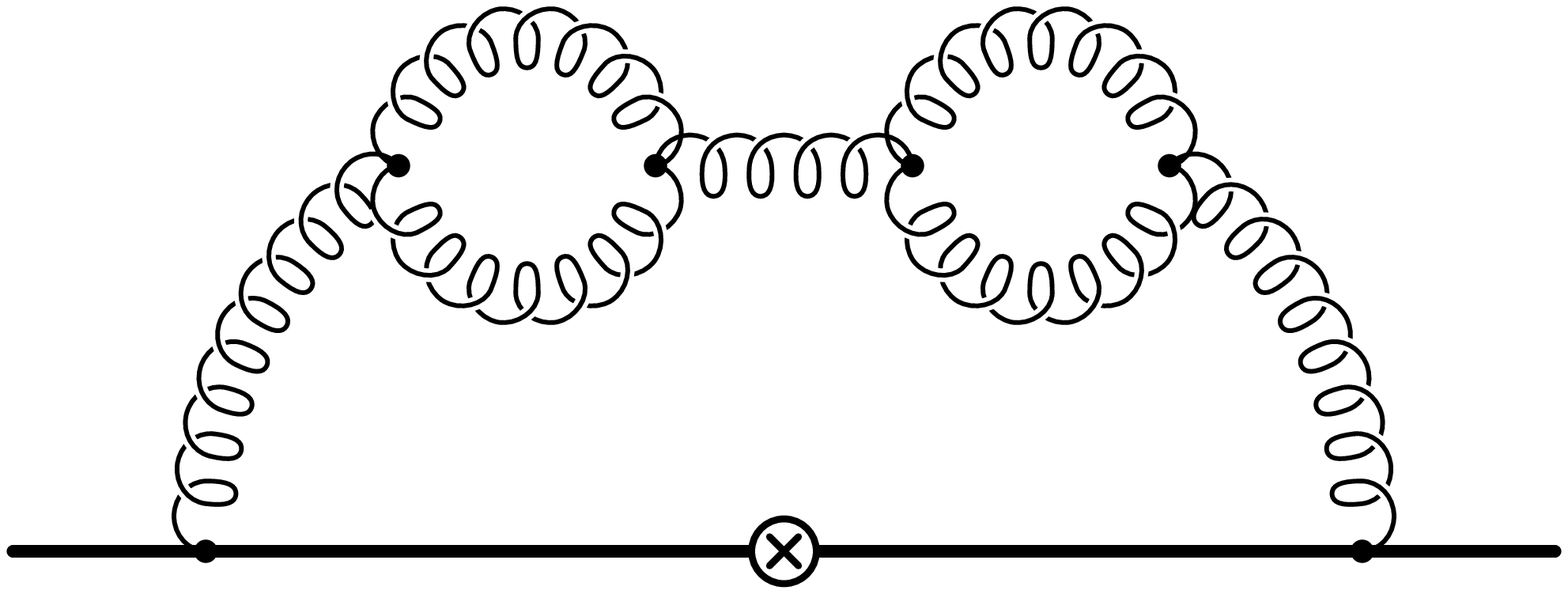,width=40mm}  &
\psfig{figure=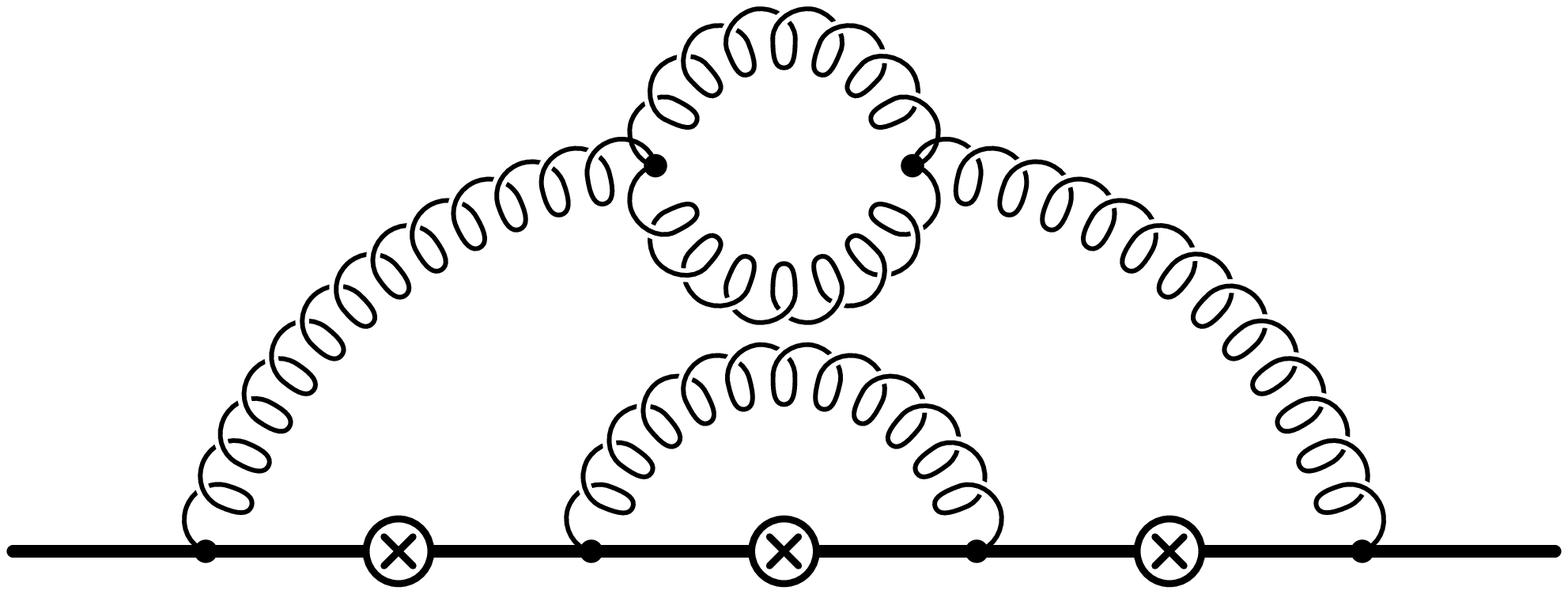,width=40mm} &
\psfig{figure=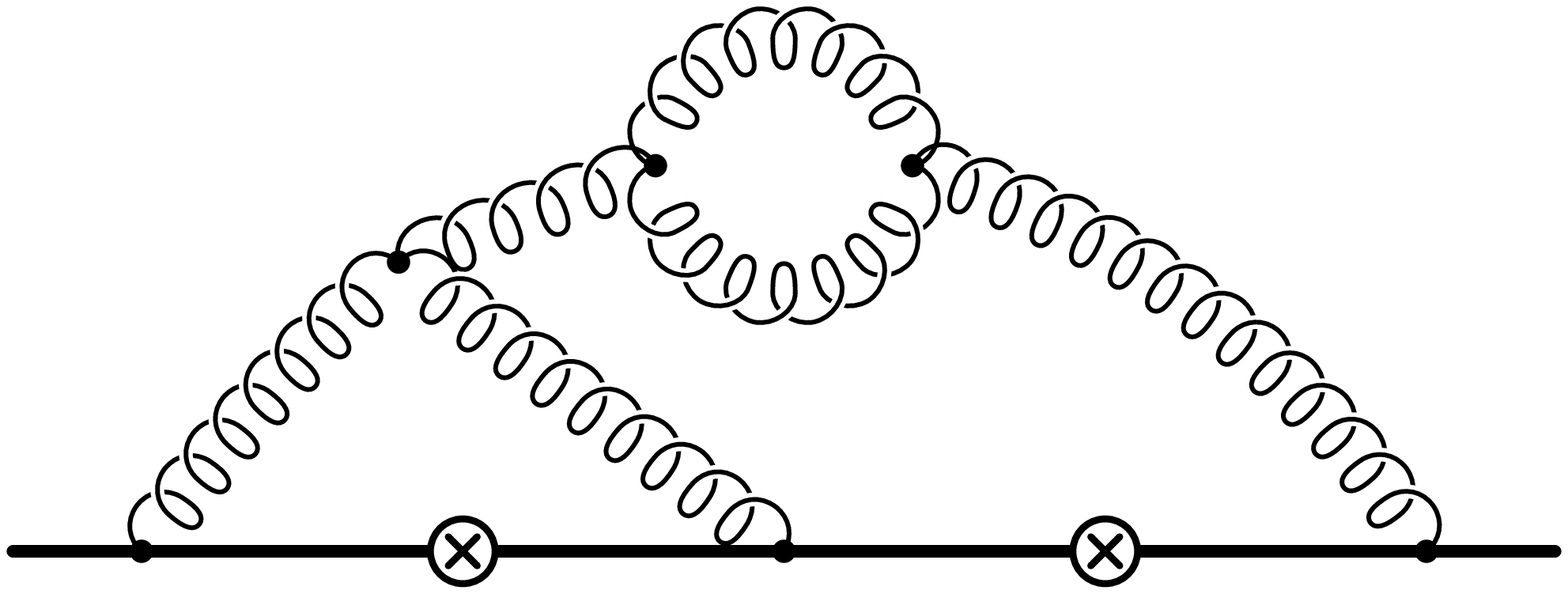,width=40mm}  \\
\psfig{figure=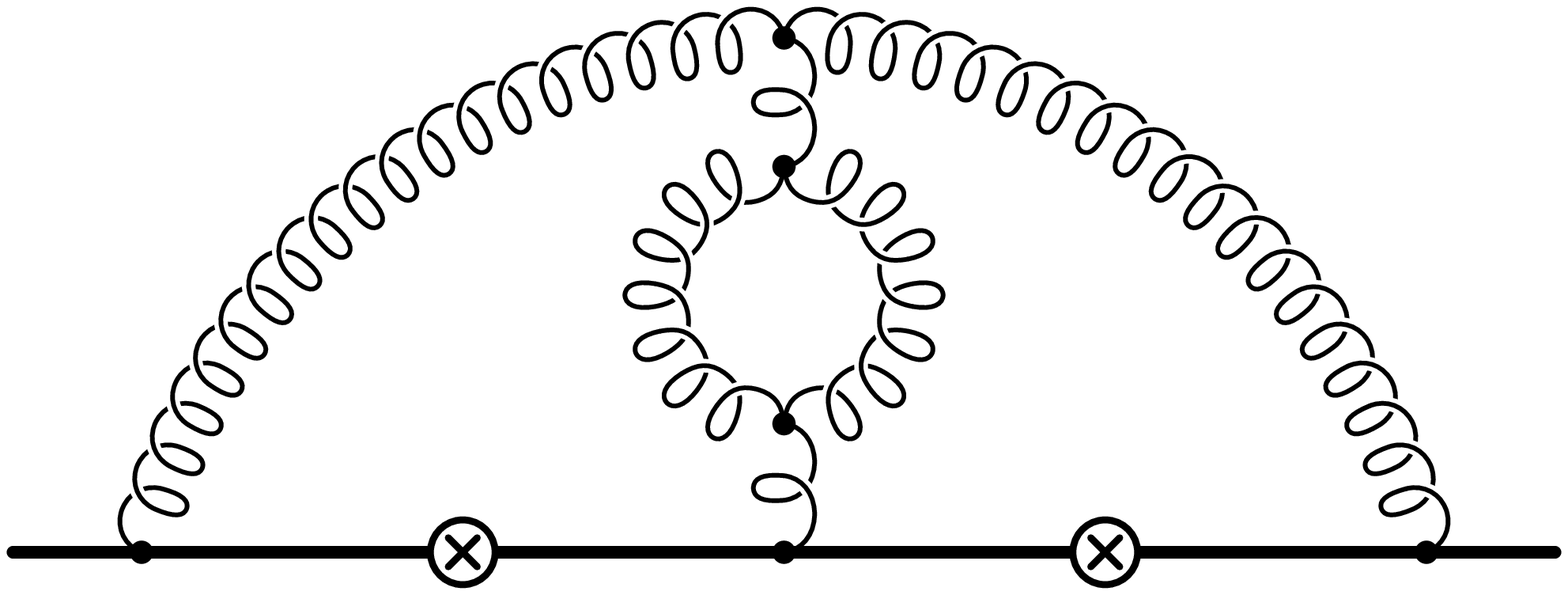,width=40mm}  &
\psfig{figure=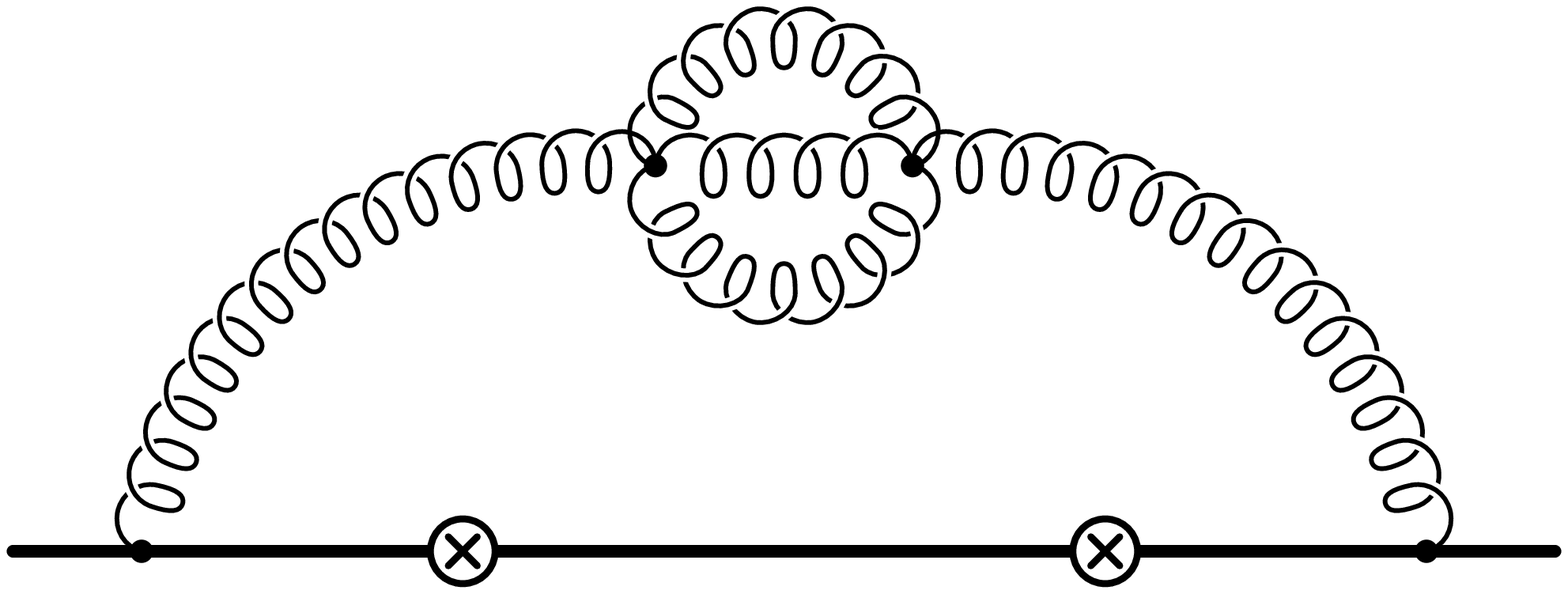,width=40mm}  &
\psfig{figure=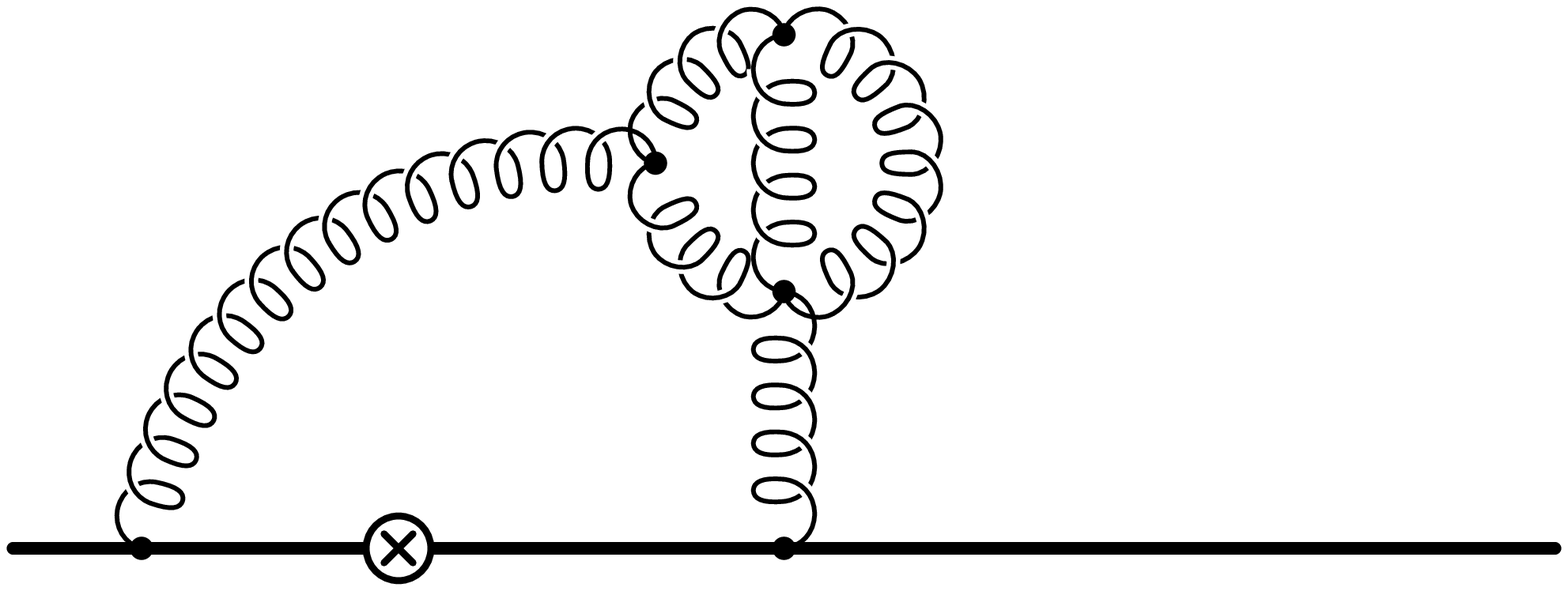,width=40mm}  \\
\psfig{figure=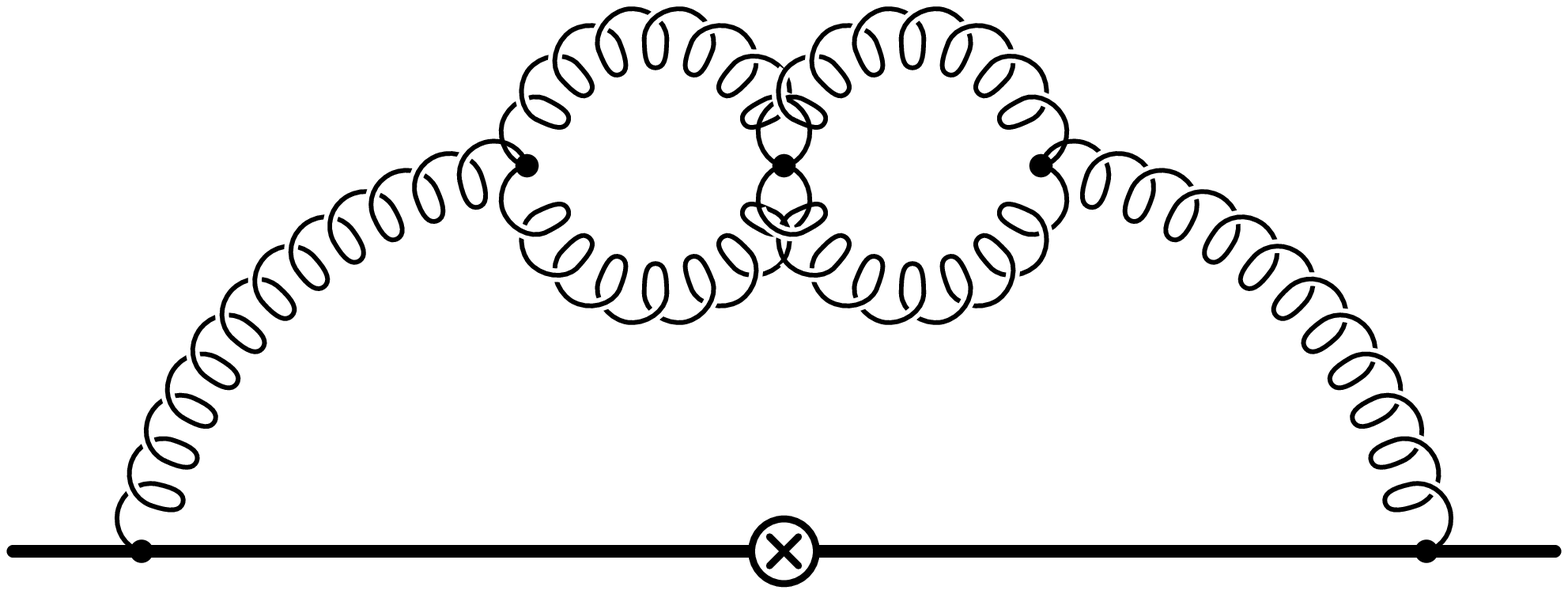,width=40mm}  &
\psfig{figure=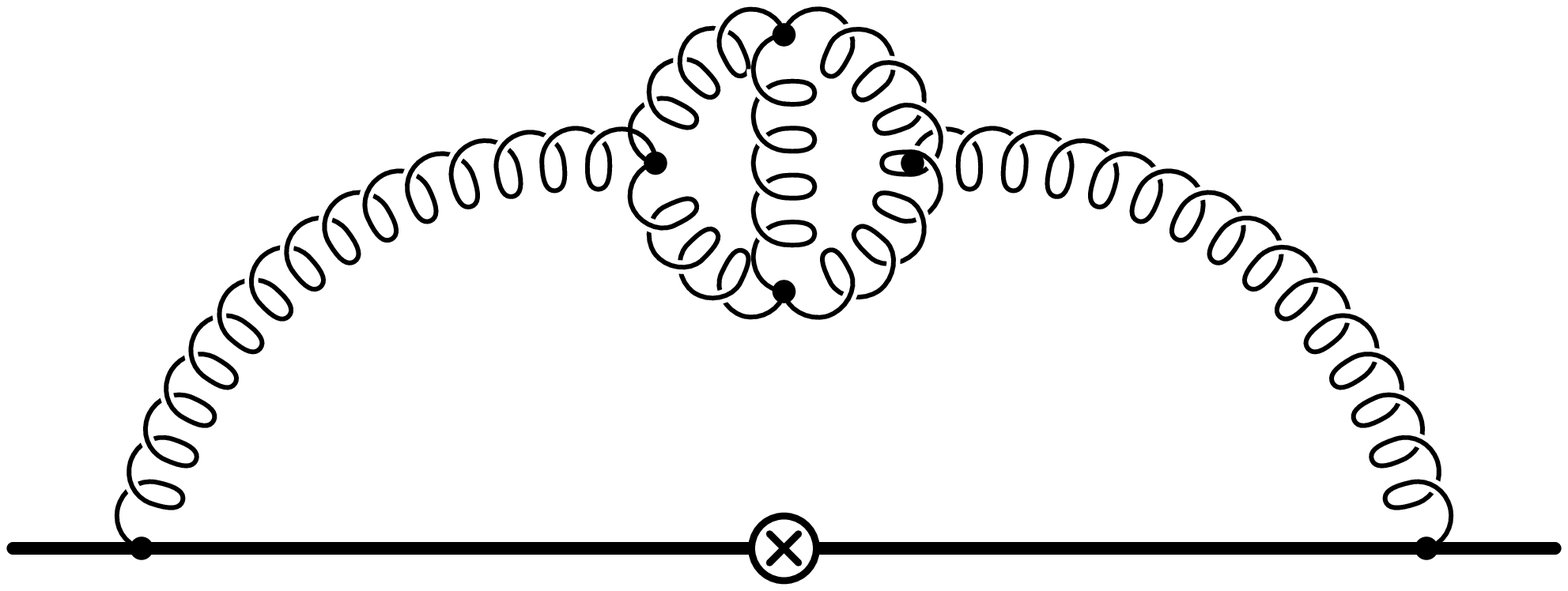,width=40mm}  &
\psfig{figure=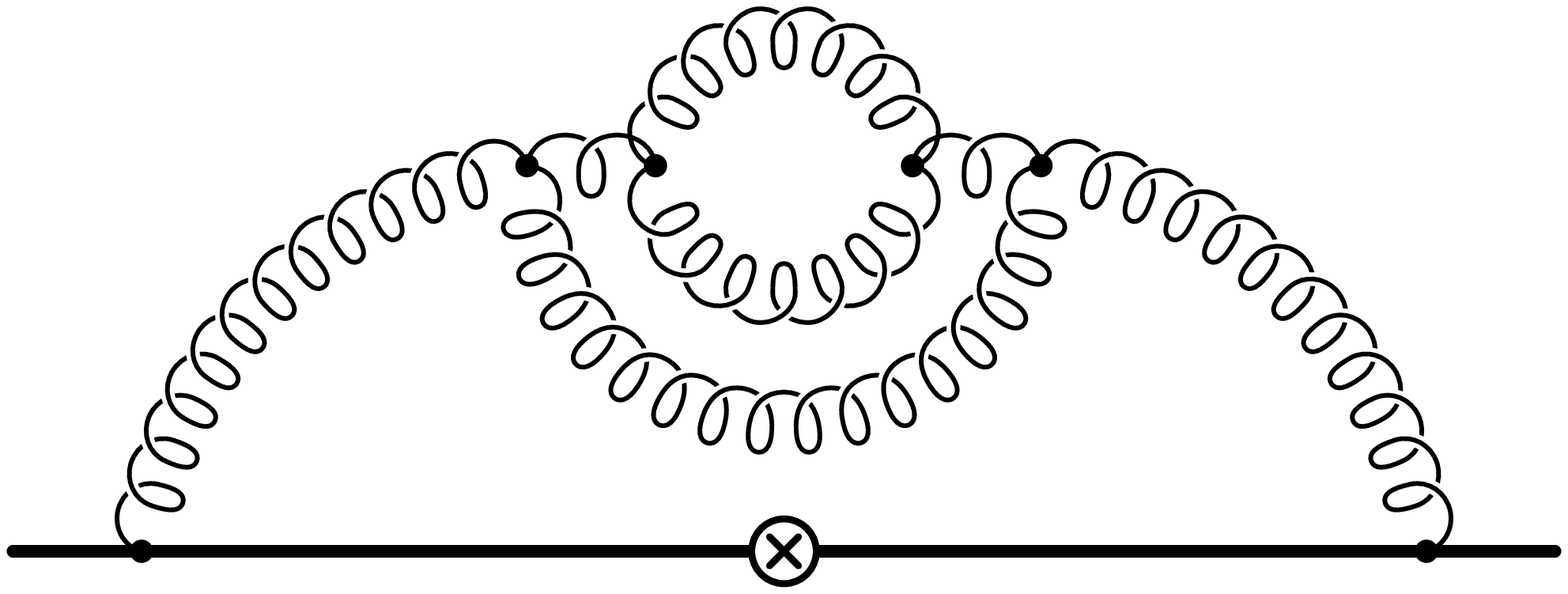,width=40mm}  \\
\psfig{figure=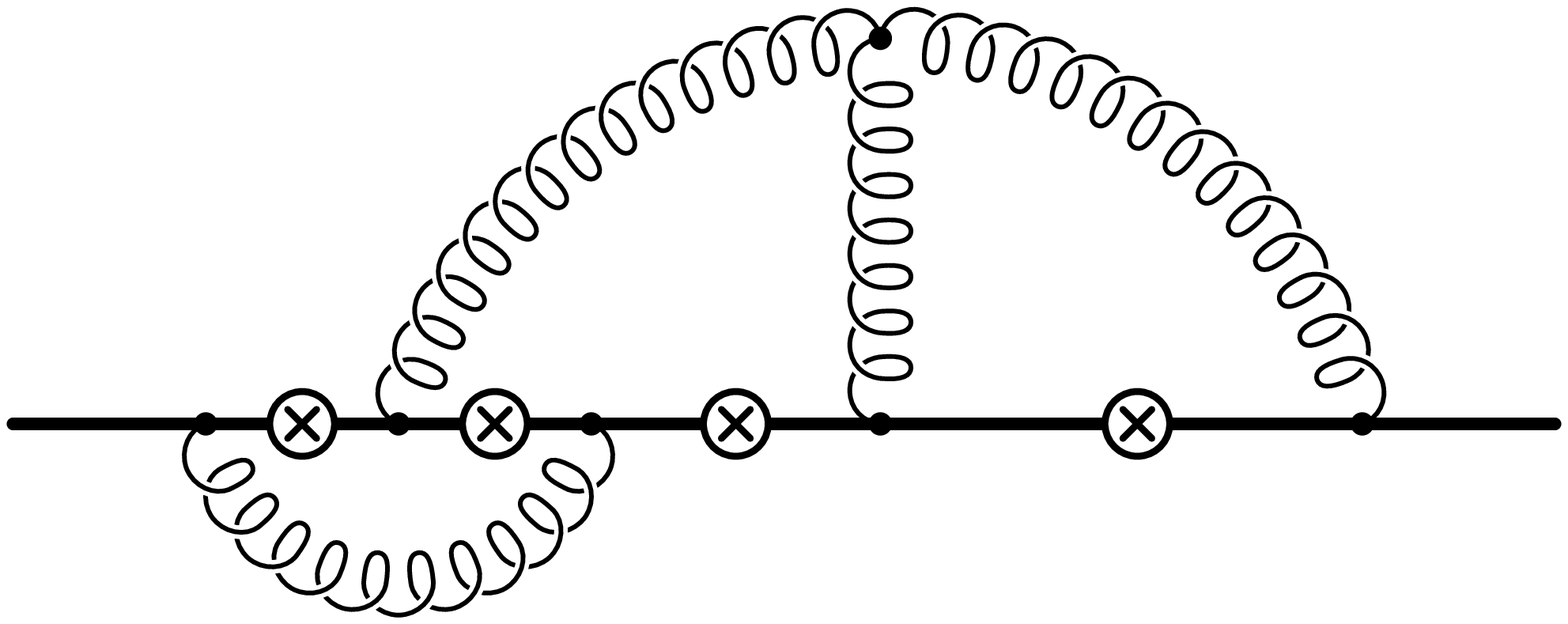,width=40mm} &
\psfig{figure=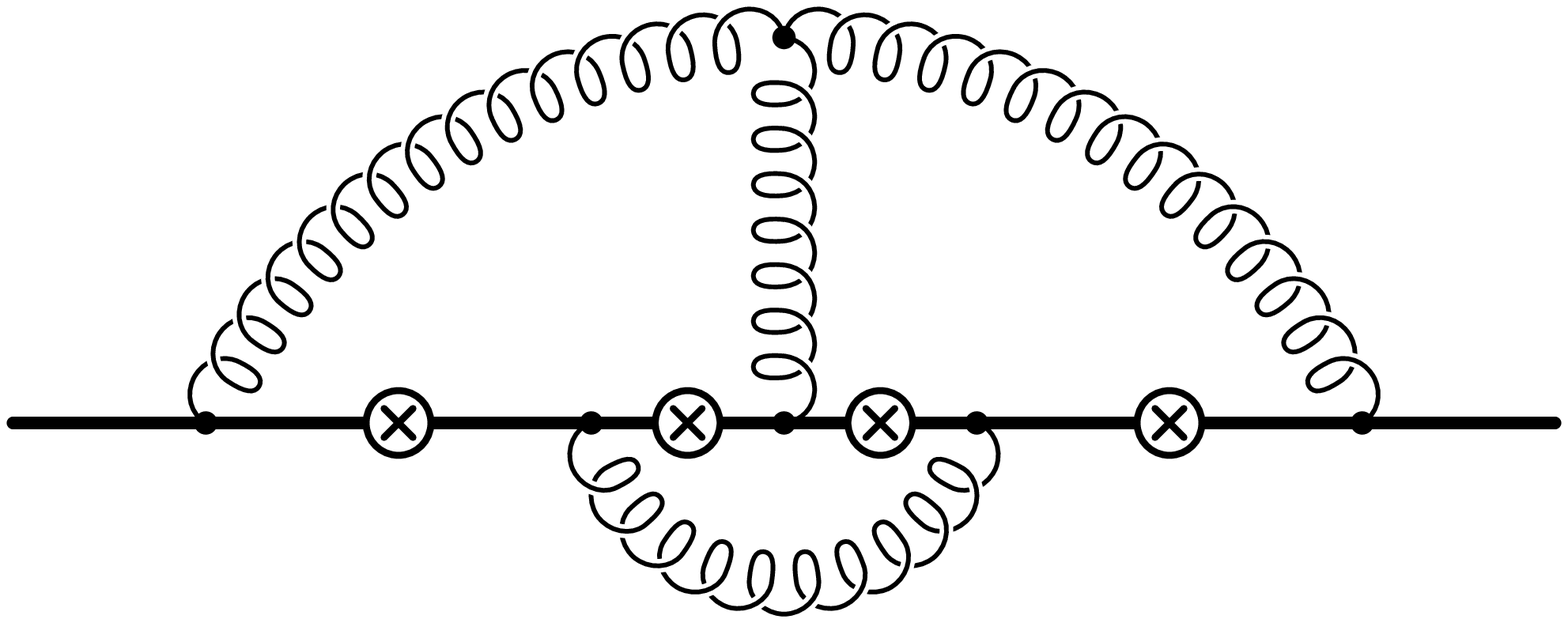,width=40mm} &
\raisebox{-1.6ex}{\psfig{figure=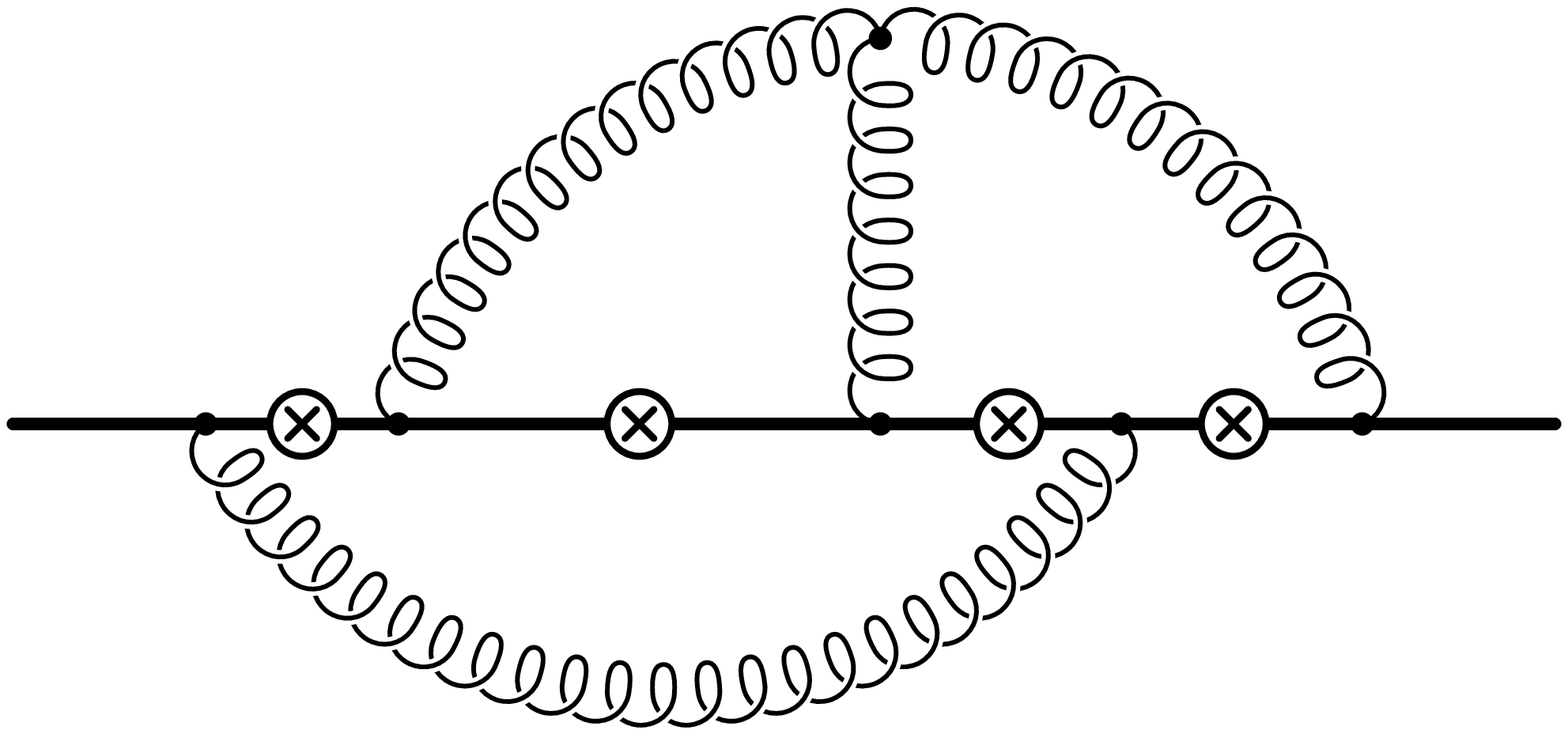,width=40mm}}
                                \\
                                &
\psfig{figure=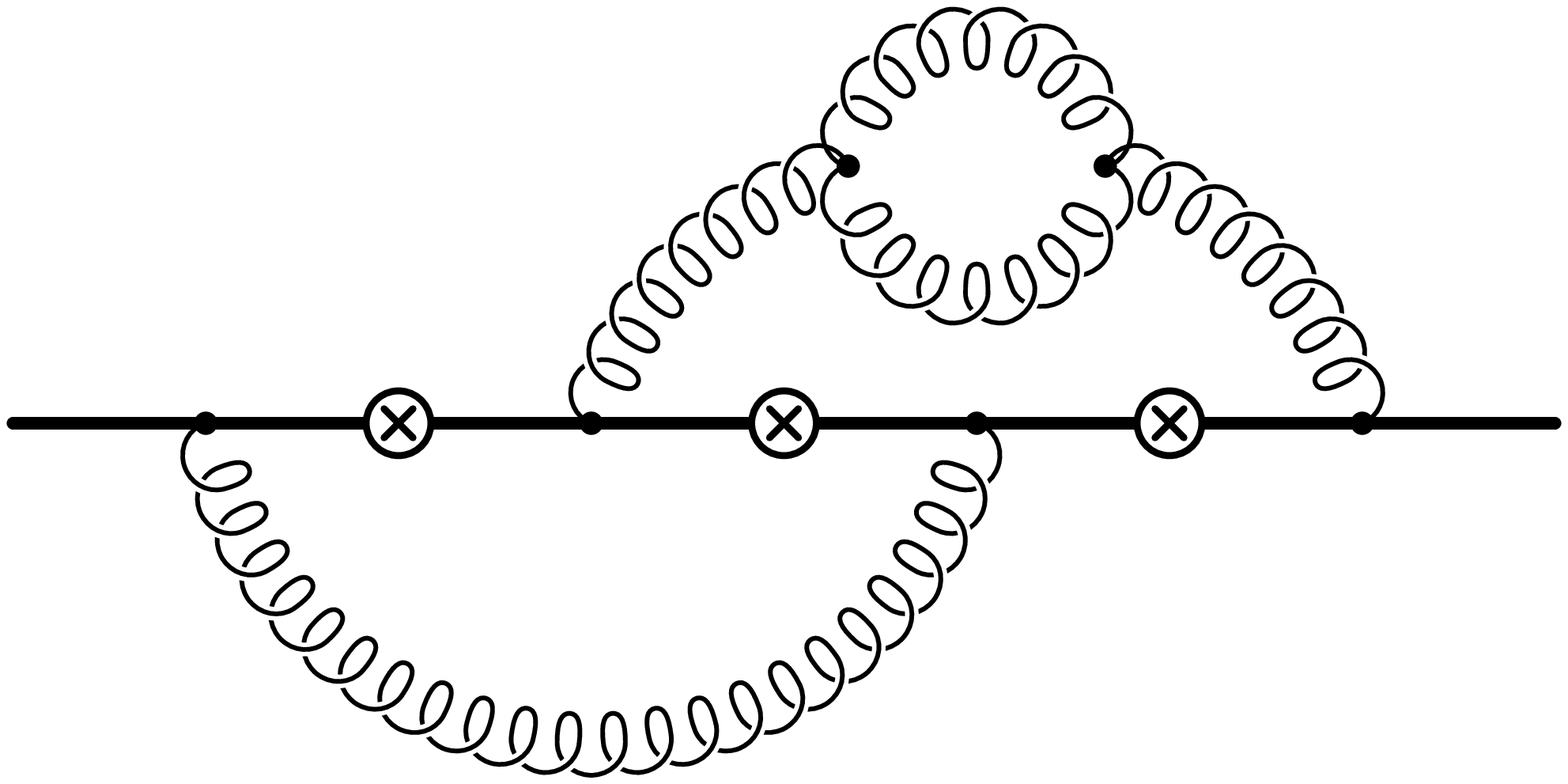,width=40mm} &
\end{tabular}
\end{minipage}
\caption{Three-loop non-abelian gluonic diagrams.} \label{fig:3loopsnab}
\end{center}
\end{figure}

The last three contributions in \eqn{eq:eta3} correspond to diagrams without closed fermion loops.  The
first and the second of them are non-abelian, and receive contributions from the diagrams shown in
Fig.~\ref{fig:3loopsnab}, as well as from the non-planar diagrams without multi-gluon vertices.  The latter,
QED-like diagrams, also give the last contribution in \eqn{eq:eta3}.

In addition to the vertex correction, the wavefunction
renormalization also  contributes to $\eta_A^{(3)}$. The
$\order{\alpha_s^3}$ part of the on-shell quark wavefunction
renormalization constant $Z_2$ was found in
\cite{Melnikov:2000zc}. To our knowledge, the present study is the
first application of that result.

One novel feature of the on-shell constant $Z_2$ at the NNNLO is
that some of the non-abelian contributions are gauge-dependent
in dimensional regularization
\cite{Melnikov:2000zc}.  In contrast, dimensionally regularized
on-shell $Z_2$ in QED is
gauge-independent to all orders, and before the explicit
calculation in \cite{Melnikov:2000zc}, it had been conjectured
that the QCD result might also be gauge-independent
\cite{Broadhurst:1991fy}.  The calculation presented here shows
that the gauge dependence of $Z_2$ found in \cite{Melnikov:2000zc}
is needed to make the full vertex correction $\eta_{A,V}^{(3)}$
gauge invariant.  The cancellation of the gauge parameter
dependence is a useful check, in addition to the vanishing of
$\eta_V^{(3)}$.

\section{Numerical results and summary}
Substituting the numerical values of the coefficients of the
SU(3) factors, we find for the first three
corrections to $\eta_A$
\ba
\eta_A^{(1)} &=& -0.5, \nn
\eta_A^{(2)} &\simeq & -0.976\, C_A  + 1.01\, C_F  - 0.0954\, N_H \, T_R  + 0.194\, N_L \, T_R , \nn
\eta_A^{(3)} &\simeq & -2.06\, C_A^2 + 0.958\, C_A \, C_F  - 2.15\, C_F^2 - 0.124\, C_A \, N_H \, T_R  +
  1.03\, C_F \, N_H \, T_R  + 2.13\, C_A \, N_L \, T_R
\nn &&  +
  0.202\, C_F \, N_L \, T_R  - 0.0507\,N_H^2\,T_R^2 -
  0.0647\, N_H \, N_L \,T_R^2 - 0.288\,N_L^2\,T_R^2.
\ea
For the case of the $b$ quark decay into a $c$ quark ($N_L=3$, $N_H=2$),
we get in the extreme zero-recoil limit,
\ba
\eta_A &\simeq &
 1
 -0.667{\alpha_s\over \pi}
 -1.85\left({\alpha_s\over \pi} \right)^2
 -11.1\left({\alpha_s\over \pi} \right)^3 + \order{\alpha_s^4}
\nn
&\simeq &
1-0.0510-0.0108-0.00495
\nn
&\simeq &
0.933.
\ea
In the last two lines we substituted $\alpha_s =0.24$, roughly (to within 20\%) corresponding to
the value appropriate for the scale $\sqrt{m_b m_c}$.  We see that the three-loop calculation presented here
contributes about 7\% of the deviation of $\eta_A$ from unity, and decreases $\eta_A$ by
about half of a percent.  A large part of this NNNLO correction
is due to the known effects that can be absorbed in the running of the coupling constant, that is
contributions of the type $\alpha_s^3 \beta_0^2$, with $\beta_0 = 11-{2\over 3}N_F$, and $N_F$ denoting the
number of fermions contributing to the running.  This is often referred to as the Brodsky-Lepage-Mackenzie (BLM)
corrections  \cite{BLM}.  We can find it using the coefficient of $N_L^2$ (diagrams containing two
light fermion loops) from \eqn{eq:eta3}.  We obtain,  taking $N_F=4$ quarks in the running,
\ba
C_F\eta_A^{(3)}({\rm BLM}) &=&
\left( 4-{33\over 2}\right)^2 C_F T_R^2    \left( {25 \over 324} - {\pi^2 \over 27}  \right)
\nn
&=& -15.0.
\ea
We see that the remaining, ``genuine'', three-loop correction is small,
\ba
C_F \eta_A^{(3)}(\mbox{non-BLM}) &=& -11.1 +15.0 = 3.9.
\label{non}
\ea

It is possible to extend this result beyond the limit of zero
momentum transfer to the leptons by expanding in the difference
of the heavy quark masses.  However, we believe that our result
estimates the true value of the correction at physical values of
$m_b$ and $m_c$  to within at most a factor of two.  The corrections due to unequal $m_b$ and $m_c$ are
suppressed by two powers of $1-m_c/m_b$ \cite{Uraltsev95}. Given that the
new non-BLM effect in \eqn{non} increases $\eta_A$ by about
$0.2\%$, we believe that it is safe to estimate the corresponding
effect for the real $b\to c$ decay at zero recoil as $0.2(2)\%$.
Finally, we note that $\eta_A$ determines corrections to the total semileptonic
decay rate if the difference of the quark masses is relatively small \cite{Shifman:1994hj} and
can be used  to estimate the $b\to c$ decay rate  \cite{Czarnecki:1998ry}.  Such analysis will be
presented elsewhere.

\emph{Acknowledgements:}  We thank Ian Blokland and Nikolai
Uraltsev for helpful discussions, reading the manuscript and
suggesting corrections. This research was supported by the Science
and Engineering Research Canada and by the Collaborative Linkage
Grant PST.CLG.977761 from the NATO Science Programme.


\end{document}